\documentclass[]{aa}
\usepackage[varg]{txfonts}
\usepackage{graphicx,grffile, booktabs}
\usepackage{amssymb,float,natbib}
\usepackage[version=4]{mhchem}
\usepackage[]{hyperref}
\usepackage{sidecap}
\hypersetup{colorlinks=true,citecolor=blue,linkcolor=black,
filecolor=black, pdfmenubar=true}
\bibpunct{(}{)}{;}{a}{}{,} 
\usepackage{xcolor}
\newcommand{\kms}{km~s$^{-1}$}
\newcommand{\heo}{\ce{H2^{18}O}}

\newcommand{\hcop}{\ce{HCO^{+}}}
\newcommand{\mco}{\ce{^{12}CO}}
\newcommand{\ceo}{\ce{C^{18}O}}
\newcommand{\tco}{\ce{^{13}CO}}

\begin{document}

   \title{Resolved molecular line observations reveal an inherited molecular layer in the young disk around TMC1A}

   \author{
   D.~Harsono\inst{1,2}
   \and
  M.~H.~D.~van der Wiel \inst{3}
   \and
   P.~Bjerkeli \inst{4}
   \and
   J.~P.~Ramsey \inst{5}
   \and
   H.~Calcutt \inst{4}
   \and \\
   L.~E.~Kristensen\inst{6}
   \and
   J.~K.~J{\o}rgensen\inst{6}
   }

    \institute{
        Leiden Observatory, Leiden University, Niels Bohrweg 2,
        2300 RA, Leiden, the Netherlands \\
        \email{dsharsono@asiaa.sinica.edu.tw}
        \and
        Institute of Astronomy and Astrophysics, Academia Sinica,
        No. 1, Sec. 4, Roosevelt Road, Taipei 10617, Taiwan, R.~O.~C.
        \and
        ASTRON, the Netherlands Institute for Radio Astronomy,
        Oude Hoogeveensedijk 4, 7991 PD Dwingeloo,
        The Netherlands
        \and
        Chalmers University of Technology, Onsala Space Observatory,
        439 92 Onsala, Sweden
        \and
        Department of Astronomy, University of Virginia, Charlottesville,
        VA 22904, USA
        \and
        Niels Bohr Institute and Centre for Star and Planet Formation,
        University of Copenhagen, {\O}ster Voldgade 5--7,
        DK-1350 Copenhagen K, Denmark
    }

    \date{Received 19 June 2020 / Accepted 23 October 2020}

  \abstract
   {
   Physical processes that govern the star and planet formation sequence influence the chemical composition and evolution of protoplanetary disks.  Recent studies allude to an early start to planet formation already during the formation of a disk.  To understand the chemical composition of protoplanets, we need to constrain the composition and structure of the disks from whence they are formed.
   }
   {
   We aim to determine the molecular abundance structure of the young disk around the TMC1A protostar on au scales in order to understand its chemical structure and any possible implications for disk formation.
   }
   {
   We present spatially resolved Atacama Large Millimeter/submillimeter Array observations of CO, HCO$^{+}$, HCN, DCN, and SO line emission, as well as dust continuum emission, in the vicinity of TMC1A.  Molecular column densities are estimated both under the assumption of optically thin emission from molecules in local thermodynamical equilibrium (LTE) as well as through more detailed non-LTE radiative transfer calculations.
   }
   {
   Resolved dust continuum emission from the disk is detected between 220 and 260 GHz.  Rotational transitions from \ce{HCO+}, \ce{HCN}, and \ce{SO} are also detected from the inner 100 au region.  We further report on upper limits to vibrational HCN $\upsilon_2 = 1$, DCN, and \ce{N2D^{+}} lines.  The \ce{HCO^{+}} emission appears to trace both the Keplerian disk and the surrounding infalling rotating envelope.  HCN emission peaks toward the outflow cavity region connected with the CO disk wind and toward the red-shifted part of the Keplerian disk.  From the derived \ce{HCO^{+}} abundance, we estimate the ionization fraction of the disk surface, and find values that imply that the accretion process is not driven by the magneto-rotational instability.  The molecular abundances averaged over the TMC1A disk are similar to its protostellar envelope and other, older Class II disks.  We meanwhile find a  discrepancy between the young disk's molecular abundances relative to Solar System objects.
}
   {
Abundance comparisons between the disk and its surrounding envelope for several molecular species reveal that the bulk of planet-forming material enters the disk unaltered.  Differences in \ce{HCN} and \ce{H2O} molecular abundances between the disk around TMC1A, Class II disks, and Solar System objects trace the chemical evolution during disk and planet formation.
   }


	\titlerunning{Resolved molecular line observations toward TMC1A}
	\authorrunning{D.~Harsono et al.}

	\maketitle


%
%
\section{Introduction}
\label{sec:intro}

The properties of newborn planets will likely reflect the environment in which they formed. Thus, knowing the physical and chemical structure of protoplanetary disks is absolutely necessary in order to understand the process of planet formation.  Evidence for on-going planet formation has been captured by recent high-spatial resolution millimeter and high-contrast IR observations that reveal gaps, rings, and spiral dust substructures in protoplanetary disks \citep[e.g.,][]{hltau2014,lperez16sci,flong18,huang18}.  The immediate emerging paradigm is that planet formation starts early during the embedded stage of star formation \citep[Class 0 and I, e.g.,][]{harsono18}.  This is supported by cosmochemical measurements of calcium aluminium inclusions and chondrules that are found in chondritic meteorites formed in the early Solar System  \citep{connelly12,bollard2017,jacquet19, haugbolle2019}.  A spatially resolved molecular studies of a young disk can reveal the environment of the early stages of planet formation.

Understanding the physical and chemical structure of planet-forming disks relies on both high-spatial resolution and spectrally resolved observations.  From dust continuum observations, the physical structure of Class II disks have been constructed and compared to theoretical studies of disk evolution \citep{wc11,manara17,tazzari17,manara19b}.  Once a physical structure is determined, the chemical structure can be studied and compared with evolutionary models coupled to chemistry \citep[e.g.,][]{dutrey97, zadelhoff01, aikawa03, ceccarelli05, pietu07, bergin10, bruderer12, chapillon12, vandermarel14, loomis15, guzman15, fedele16, salinas16, kama16, mcclure16, oberg17, carney18}.   Meanwhile, on-going planet formation can also affect the strength of molecular line emission, as suggested by a decreasing \ce{HCN}/\ce{H2O} mid-IR line flux ratio with decreasing disk mass observed toward Class II disks \citep{najita13}.  However, similar studies towards embedded, young stellar objects are still difficult.

Detailed studies of the physical and chemical structure of young, embedded disks are hindered by the surrounding, obscuring protostellar envelope.  In addition, since the chemical evolutionary timescale in the disk is typically longer than the envelope infall timescale, the final chemical structure of the disk also depends on the physical and chemical evolution of the protostellar envelope  \citep{bergin97,aikawa99,visser09,hincelin13,yoneda16}.  In this regard, chemical studies with single-dish telescopes have been instrumental in determining the bulk chemical structure of large-scale envelopes around low-mass stars \citep[$> 1000$ au, e.g.,][]{blake95,vandishoeck98,jorgensen04c, graninger16} down to the scales of a few hundred au with millimeter interferometers such as the Submillimeter Array and IRAM NOrthern Extended Millimeter Array \citep[e.g.,][]{jorgensen05b,prosac07,bisschop08,maury14,taquet15}.   With the Atacama Large Millimeter/submillimeter Array, it is now possible to spatially and spectrally resolve the molecular emission to isolate the Keplerian disk from the surrounding envelope \citep[e.g.,][]{sakai14, delavillarmois18, delavillarmois19b}.  This aspect makes it possible to explore the chemical structure of young disks and examine the early stages of planet formation.

Many inferred chemical compositions demonstrate a link between Solar System comets and young stellar objects, while others indicate differences since formation \citep[e.g.,][]{schoier02,visser09, pontoppidan14,persson14, garrod19}.  For example, the most recent comparison between the young stellar object IRAS 16293--2422 and comet 67P/Churyumov--Gerasimenko shows a correlation in the CHO-, N- and S-bearing molecules \citep{drozdovskaya19}.  Meanwhile molecular species such as \ce{CH3OH}, \ce{CH3CN}, and \ce{CH3SH} are elevated compared to comets, which indicates some degree of processing.   Further evidence for a direct link between the birth environment of a planet and the initial structure of the proto-Solar disk can be found in the elemental enhancement of Jupiter's atmosphere with respect to solar \citep{owen99,asplund09, oberg19, bosman19}.  It is therefore crucial to understand the chemistry and variation of molecular abundances during the early stages of disk evolution to be able to draw conclusions about the details of planetary composition.

In the classical theory of a disk \citep{lodato08, armitage11}, a protoplanetary disk is characterized by its Keplerian motion.  Hence, molecular emission from rotationally supported object will more likely trace the chemical structure of an early disk.  A few young stellar objects show a clear sign of an embedded Keplerian disk \citep[e.g.][]{prosac09, harsono14, yen17, delavillarmois19a}.  Only a handful of these targets are less embedded and at a favorable orientation such that there is a direct line-of-sight to the disk through the tenuous envelope.  One of those targets is TMC1A (IRAS 04365+2535) with $i \sim 60^{\circ}$ and a $\sim$0.1 $M_{\odot}$ envelope, which makes it a ``Stage I'' embedded young stellar object \citep[][]{robitaille06,kristensen12}.  A ``Stage I'' system refers to a protostellar object with an envelope mass that is similar to its protostellar mass.   High-spatial resolution observations of \mco, \tco, and \ceo\ demonstrate a clear distinction between outflowing gas \citep{bjerkeli2016}, the Keplerian disk, and the large-scale envelope \citep{harsono18}.  The relatively simple geometry of TMC1A with its well-defined disk, makes it an ideal laboratory for determining the distribution and abundances of common, simple molecules on scales of $\sim 15$ au. This paper presents such an analysis and compares the inferred chemistry to that of the Solar System as well as more evolved young stellar objects. Thereby it also helps shedding light on which molecules can serve as fingerprints of  disk formation and evolution.

This paper presents the detection of simple molecules in the  young disk around TMC1A.  By determining their molecular column densities, we constrain the molecular abundance structure of its Keplerian disk with respect to \ceo.  The paper outline is as follows.  Section \ref{sec:obsdetails} presents the observational details.  Dust continuum emission and the detected molecular lines are presented in Sect.~\ref{sec:obsresults}.  HCN and DCN lines are analyzed further using the \ceo\ spectral cube as a proxy mask.  In Sect.~\ref{sec:colden_excitation}, optically thin, thermalized molecular emission and non-LTE radiative transfer calculations are been used to determine the molecular column densities and their excitation conditions.  Using these column densities, we estimate the molecular abundances and temperature structure of the young disk around TMC1A.  By utilizing a reduced chemical network, the ionization fraction of the disk is approximated in order to understand the physical processes that drive its accretion.  The derived abundances are compared with the protostellar envelope, Class II protoplanetary disks, and Solar System objects in Sect.~\ref{sec:molecularlayer}.  Finally, the summary and conclusions can be found in Sect.~\ref{sec:summaryconclude}.

%
%

\begin{table*}
    \caption{ALMA observational details on Band 6 long baseline
    observations of TMC1A.  Project IDs are indicated. }
    \label{tbl:obsdetails}
    \centering
    \begin{tabular}{c c c c c c c c}
        \hline \hline
        Date    &   Bandpass    &   Phase   & Flux   &
            Bl. coverage (km)  & $N_{\rm ant.}$ & PWV (mm) & WVR
            corr. \tablefootmark{a}\\
        \hline
        \multicolumn{8}{c}{2015.1.01549.S}
        \\
        Oct 16, 2015    &  J0510+1800   &  J0440+2728   & J0433+2905    &
            0.04 -- 16.2      & 42        & 1.8   		& Yes
        \\
        Oct 23, 2015    &  J0510+1800   & J0440+2728    & J0433+2905    &
            0.04 -- 16.2      & 40        & 0.6  	& Yes
        \\
        Oct 30, 2015    & J0510+1800   & J0440+2728    & J0433+2905    &
            0.04 -- 16.2      & 40        & 0.3  		& Yes
        \\
        \hline
        \multicolumn{8}{c}{2016.1.00711.S}
        \\
        Sep 22, 2017    & J0510+1800    & J0438+3004    & J0510+1800    &
            0.04 -- 12.1    & 42    & 0.7       & No
        \\
        Sep 23, 2017    & J0510+1800    & J0438+3004    & J0510+1800    &
            0.04 -- 12.1    & 40    & 0.4       & No
        \\
        \hline
        \multicolumn{8}{c}{2017.1.00212.S}
        \\
        Oct 19, 2017    & J0510+1800    & J0438+3004    & J0510+1800    &
            0.04 -- 16.2    & 51    & 1.1       & No
        \\
        \hline \hline
    \end{tabular}
\tablefoot{
  \tablefoottext{a}{Application of the non-standard WVR correction as presented in \citet{maud17wvr}.
    }
  }
\end{table*}

\section{Observations}
\label{sec:obsdetails}

TMC1A was observed during long baselines campaigns (LBC, 16 km) of the Atacama Large Millimeter/submillimeter Array (ALMA) on three occasions.  We have concatenated three LBC projects: 2015.1.01549.S (PI: van der Wiel), 2016.1.00711.S (PI: van der Wiel), and 2017.1.00212.S (PI: Bjerkeli).  The first data set (2015.1.01549.S, tuned to frequencies of CO and its isotopologs) was published in \citet{bjerkeli2016} and \citet{harsono18}.  We also use water (\ce{H2^{18}O} $3_{1,3} - 2_{2,0}$ at 203 GHz) observations taken with the NOrthern Extended Millimeter Array (NOEMA), and which are presented in \citet{harsono20}.  Here, we present observational details of the second (2016.1.00711.S, Sect.~\ref{sub:obs_hcn}) and third (2017.1.00212.S, Sect.~\ref{sub:obs_continuum}) ALMA data sets, while we refer to \citet{harsono18,harsono20} for details on the imaging and calibration of the first ALMA data set and the NOEMA data, respectively.  Table~\ref{tbl:obsdetails} lists the observational details of the ALMA programs used in this paper.

\subsection{ALMA observations: 2016.1.00711.S}
\label{sub:obs_hcn}

ALMA observed TMC1A on September 22 and 23 in 2017.  The observations were taken with $\sim$40 antennas under very good weather conditions with a precipitable water vapor (PWV) of 0.35 mm.  The baseline coverage was between 40 m and 12 km, which translates to 30 k$\lambda$ up to 10$^{4}$ k$\lambda$.  The bandpass, phase, and flux calibrators are indicated in Table~\ref{tbl:obsdetails}.  Four spectral windows were utilized in this program.  Three narrow high resolution windows were centered on HCN $J=3-2$ (265.8861800 GHz), HCO$^{+}$ $J=3-2$ (267.8527094 GHz), and SO $N_{J} = 5_6 - 4_5$ (251.82577000 GHz), respectively.  A wide spectral window (1.8 GHz) was centered at 253.2 GHz to measure the dust continuum.  Rest line frequencies are obtained from the JPL  \citep{JPLdb} and CDMS catalogs \citep{cdms1,cdms2}.  The data were calibrated using the Cycle 4 pipeline of \textsc{CASA} 4.7.2 \citep{CASA}.

\subsection{ALMA observations: 2017.1.00212.S}
\label{sub:obs_continuum}

This ALMA project was executed on October 19 2017.  The observations were taken with $\sim$51 antennas.  The baseline coverage was between 40 m and 16.2 km, which translates to 30 k$\lambda$ to $10^{4}$ k$\lambda$.  The bandpass, phase, and flux calibrators are also indicated in Table~\ref{tbl:obsdetails}.  Four broadband continuum spectral windows (1.8 GHz) were used in this program.  Molecular lines are not clearly detected because of the low spectral resolution. The data was calibrated using the Cycle 5 pipeline of \textsc{CASA} 5.1.1.

\subsection{Self-calibration and combined ALMA data}
\label{sub:selfcalcombined}

\begin{table*}
    \caption{
    Millimeter interferometric observations presented in this paper.  Synthesized beams and noise levels of the images are listed below.  Upper limits are calculated over the size of the dust continuum emission.  We report the noise to be 10\% of the integrated flux density unless the measured noise is larger than 10\%. The noise level for the molecular line observations is per velocity channel (0.3 km s$^{-1}$).}
    \label{tbl:imgs}
    \centering
    \begin{tabular}{cccccccc}
        \hline \hline
        Name    &   Frequency   & $E_{\rm up}$
        & $\log_{10} A_{\rm ul}$  & Beam\tablefootmark{a} &  Noise
        & Int. time & Integrated flux density
        \\
                    & (GHz)             & (K)     &  (s$^{-1}$)
                & ($'' \times '', ^{\circ}$)      & (mJy bm$^{-1}$)
                & (min.)     & (mJy / mJy km s$^{-1}$)
        \\
        \hline
        NOEMA Continuum   & 203      & ... & ...  & $0.78 \times 0.72, 62$
        & 1.1      & 180     & $210\pm20$
        \\
        NOEMA Continuum   & 220     & ... & ...  & $0.79 \times 0.61, 54$
        & 1.2      & 180     & $170 \pm 17$
        \\
        ALMA Continuum   & 220      & ... & ... & $0.031\times 0.020, -3$   &
            0.08       & 170     & $220 \pm 22$
        \\
        ALMA Continuum   & 230      & ... & ... & $0.032\times 0.021, 1$   &
            0.05       & 111    & $240 \pm 24$
        \\
        ALMA Continuum   & 240      & ... & ... & $0.028\times 0.018, 0$   &
            0.08       & 90    & $280 \pm 28$
        \\
        ALMA Continuum   & 260      & ... & ... & $0.049\times 0.032, 11$   &
            0.05       & 63     & $190 \pm 19$
        \\
       \heo\ $3_{1,3} - 2_{2,0}$           & 203.4075
        &  203.7       						  	    & $-5.32$
        & $0.78 \times 0.72$, 59			    & 8
        & 180                                           & $< 11$
        \\
        \ce{DCN} 3--2                              &  217.2385
        &  20.85        						        & $-3.34$
        &  $0.13 \times 0.10$, 32	        & 1.5             & 91
        & $ < 8$
        \\
        \ce{C^{18}O} 2--1                    &  219.5604
        &   15.80       						        & $-6.22$
        &  $0.13 \times 0.10$, 31 	        & 1.8             & 91
        & $330 \pm 33$
        \\
        \ce{^{13}CO} 2--1                        &   220.3987
        &   15.87                                       & $-6.22$
        &  $0.13 \times 0.10$, 32		    & 2.4         & 85
        & $290 \pm 30$
        \\
        \ce{^{12}CO} 2--1                    &   230.5380
        &   16.60       							    &   $-6.16$
        &  $0.12 \times 0.10$, 0.5 		    & 2.5         & 83
        & $1740 \pm 174$
        \\
        \ce{N2D+} $J = 3 -2$             &    231.3216
        &  22.20                                     &  $-2.67$
        & $0.12 \times 0.10$, 32         & 0.8     &  91
        & $100 \pm 22$
        \\
        \ce{SO} $N=6_{5}-5_{4}$        & 251.8258
            & 50.66                                     & $-3.71$
            & $ 0.13 \times 0.11$, 0.1         &  1.0        & 71
            & $460 \pm 46$
        \\
        \ce{HCN} $v_2 = 1$ $J=3-2e$         &  265.8527
        &  1050         							        & $-2.64$
        &  $0.092 \times 0.07$, 0.4		        & 0.9         & 71
        & $< 20$
        \\
        \ce{HCN} $3-2$                      &  265.8862
        &   25.52       						            & $-3.55$
        &  $0.12 \times 0.10$, 0.1	            & 1.3             & 71
        & $160 \pm 20$
        \\
        \ce{HCN} $v_2 = 1$ $J=3-2f$       &  267.1993
        &  1050         							        & $-2.63$
        &  $0.092 \times 0.07$, 20	            & 0.9         & 71
        & $< 20$
        \\
        \ce{HCO+}  $J=3-2$                      & 267.5576
        & 25.68                                             & $-2.84$
        & $0.12 \times 0.11$, 3.5                 & 1.2     & 71
        & $640 \pm 64$
        \\
        \hline
    \end{tabular}
\tablefoot{
  \tablefoottext{a}{Elliptical synthesized beam parametrized by the full-width half-maximum  (FWHM) of the major axis $\times$ the FWHM of the minor axis, and a position angle. }
    }
\end{table*}

We combined these ALMA projects into a single measurement set using the CASA task \textsc{concat} to improve the $S/N$ of the dust continuum model.  The concatenated data allows for simultaneous phase and amplitude self-calibrations to produce better flux calibration across the three data sets.  This self-calibration is based on the line-free channels in each spectral window, and was performed with \textsc{CASA} 5.4.1.  We fixed the phase center of each measurement to the value obtained in \citet{harsono18} utilizing the data with the most stable weather conditions.  The phase center of J2000 04h39m35.203s +25d41m44.21s is determined by fitting an elliptical Gaussian to the continuum visibilities with \textsc{CASA} task \textsc{uvfit}.  The phase solutions obtained from the broadband windows are applied across the narrow spectral windows.  The imaging of the continuum is performed with \textsc{tclean} using Briggs weighting \citep{briggs95} to minimize the side lobes.  Spectral windows containing the targeted molecular lines are continuum subtracted in the (u,v) plane with the task \textsc{uvcontsub}.

The targeted molecular lines are listed in Table~\ref{tbl:imgs} along with the noise level per spectral resolution element. The spectral lines are imaged at 0.3 km s$^{-1}$ velocity resolution including a spatial tapering at 0$\farcs$08, resulting in the synthesized beam sizes indicated in Table~\ref{tbl:imgs}.  Appendix~\ref{app:undetected} lists a few molecular lines that were present in our spectral set up, but not detected in our data sets.

%
%
\section{Observational results}
\label{sec:obsresults}

\begin{figure*}[hbt!]
 \centering
 \includegraphics[width=0.95\textwidth]{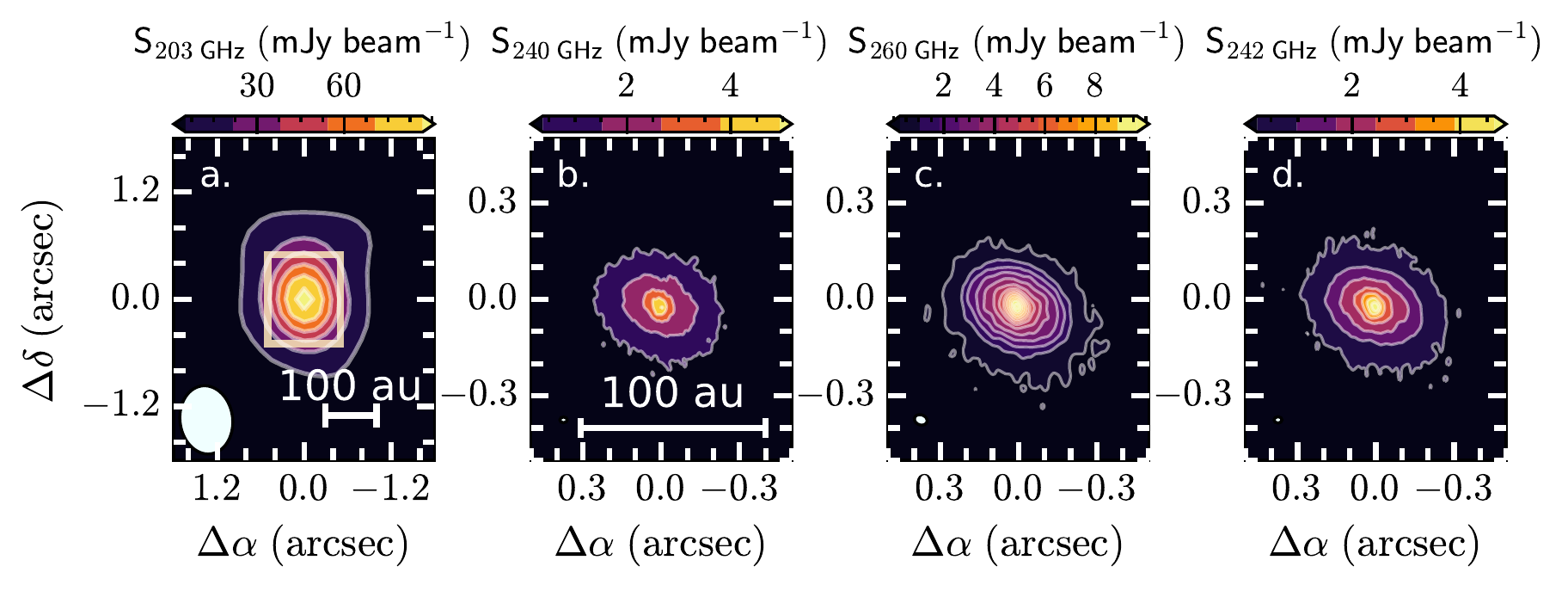}
 \caption{Dust continuum images at various frequencies. The color scale is shown on the top of each panel starting at 5$\sigma$.  The white contours indicate the 5, 20, 35,... $\sigma$ contours up to the peak intensity. A scale of 100 au is shown in the first two panels for reference. The synthesized beam is indicated by a white ellipse in the bottom left corner of each panel.  {\it a}: NOEMA observations at 203 GHz. The color scale spans linearly from 0 to the maximum intensity of 93 mJy beam$^{-1}$.  The square box indicates the $1'' \times 1''$ region of the ALMA data that is shown in panels {\it b, c} and {\it d}.  {\it b}: Dust continuum image of the 240 GHz ALMA data.  The color scale spans up to the maximum intensity of 5 mJy beam$^{-1}$.  {\it c}: Dust continuum image of the 260 GHz ALMA data.  The color scale spans from 0 to the maximum intensity of 9.8 mJy beam$^{-1}$.  {\it d}: Dust continuum emission of the combined ALMA data imaged at 242 GHz.  Similarly, the color scale spans from 0 to the maximum intensity of 5.1 mJy beam$^{-1}$.
 }
 \label{fig:contimages}
\end{figure*}

\subsection{Spatially resolved dust continuum}
\label{sub:dustcontinuum}

\begin{table}
    \caption{Best-fit parameters for the Gaussian intensity profile model.  The stated uncertainties are the 1$\sigma$ deviation from the mean of the last 1000 steps of the MCMC fit with walkers whose acceptance fraction is $>$ 20\% (see text). }
    \label{tbl:fitdust}
    \centering
    \tiny
    \begin{tabular}{cccccc}
        \hline \hline
        Name        &  Size & $i$ & $PA$ & Offset & Flux density
        \\
                        & ($''$)  & ($^{\circ}$) & ($^{\circ}$) & ($''$,$''$)
                        & (mJy)
        \\
        \hline
        NOEMA 203    & ...      & ...     & ...  & ... & ...
        \\
        NOEMA 220    & $0.2 \pm 0.1$     & $45^{+26}_{-30}$ &
            $80^{+60}_{-50}$     & 0,-0.13     & $250^{+2100}_{-220}$
        \\
        ALMA  220      & $0.1 \pm 0.01$    & $52\pm 3$     &
            $76\pm 5$    & 0.01,-0.02    & $300^{+18}_{-16}$
        \\
        ALMA  230     & $0.1 \pm 0.01$    & $51\pm 3$     &
            $75\pm 3$    & 0.01,0.001     & $310^{+12}_{-11}$
        \\
        ALMA 240      & $0.1 \pm 0.01$     & $49 \pm 3$     &
            $76 \pm 3$     & 0.01,-0.02     & $350^{+13}_{-12}$
        \\
        ALMA 260      & $0.1 \pm 0.01$     & $49 \pm 3$     &
            $76 \pm 6$    & 0.004,-0.02    & $302^{+17}_{-16}$
       \\
        \hline
    \end{tabular}
\end{table}

The continuum images of TMC1A between 203 and 260 GHz from the NOEMA and ALMA observations are shown in Fig.~\ref{fig:contimages}.  The NOEMA observation does not spatially resolve the disk, which is evidenced by the lack of elongation present in the ALMA images.  However, the deconvolved size ($0\farcs56\times 0\farcs44$), as determined by fitting an elliptical Gaussian to the observed visibilities, indicates that the dust continuum is mostly tracing the Keplerian disk ($\sim\, 100$ au, \citealt{harsono14}).  The peak intensities of the different continuum images are 93 mJy beam$^{-1}$ at 203 GHz, 5.1 mJy beam$^{-1}$ at 242 GHz, 5 mJy beam$^{-1}$ at 240 GHz, and 9.8 mJy beam$^{-1}$ at 260 GHz.

\begin{figure}
 \centering
 \includegraphics[width=0.49\textwidth]{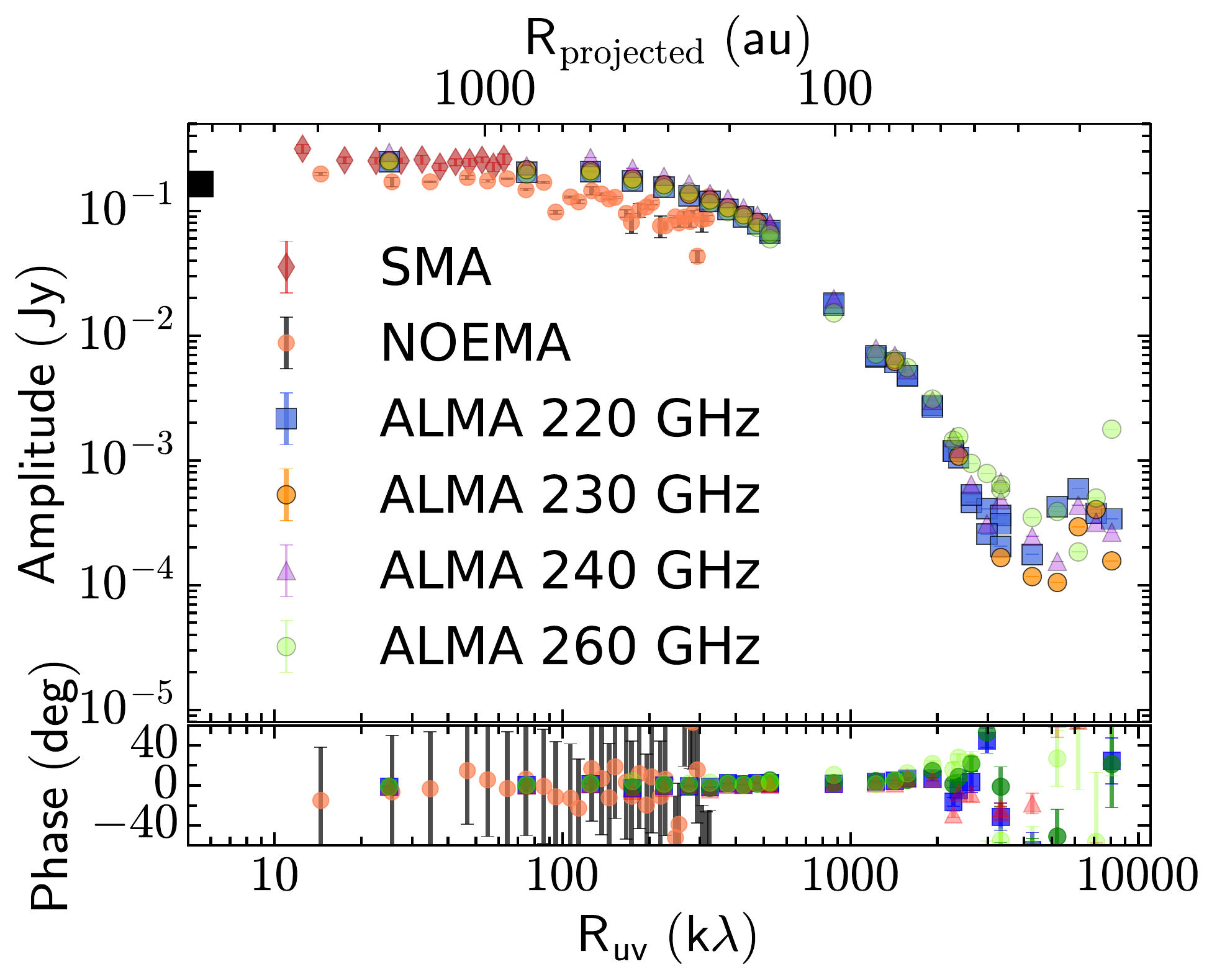}
 \caption{
 Dust continuum amplitudes and phases as a function of projected baselines $R_{\rm uv}$ in $k\lambda$.  Different colors and symbols show the various data sets used.  The black, square box near the axis shows the 850 $\mu$m single-dish flux density scaled to 1.1 mm (see \citealt{harsono14}).  The long baseline ALMA data are split into four different frequencies: 220 GHz, 230 GHz, 240 GHz and 260 GHz (see Table \ref{tbl:imgs}).  The standard deviation of each radial bin is also shown, while the corresponding physical scale in au is indicated on the top axis.}
 \label{fig:contdata}
\end{figure}

Continuum visibilities at various frequencies are shown in Fig.~\ref{fig:contdata} including the Sub-millimeter Array (SMA) data from \citet{prosac09} and NOEMA data from \citet{harsono14}.  The binned SMA data are scaled to 1.3 mm using a frequency dependence of $S_{\nu} \propto \nu^{2.5}$.  The phase as a function of projected baselines is close to zero for the ALMA observations, indicating the high data quality.  The large standard deviation in the NOEMA data reflects the lower number of (u,v) points relative to the ALMA data.  From the comparison between the integrated flux density of the NOEMA and ALMA data at 220 GHz, the uncertainty in the absolute flux density is within 10\%.  It is interesting to note that the amplitude of the 260 GHz observations is lower than the values at 220 GHz at short baselines ($< 1000 \ {\rm k} \lambda$) while the opposite is true at long baselines ($> 1000 \ {\rm k} \lambda$).  The peak intensity of the dust continuum image at 260 GHz is indeed higher than the peak at the lower frequencies.  However, as indicated in Table \ref{tbl:imgs}, the integrated continuum flux density at 260 GHz is lower than at 203 GHz in the image plane.  The lower flux density at 260 GHz may indicate resolved-out continuum emission at the higher frequency.

To characterize the dust disk, we utilize \textsc{Galario} \citep{galario}; \textsc{Galario} calculates the synthetic (u,v) visibilities given an intensity profile and makes it possible to fit Gaussian intensity profiles to the observed visibilities while simultaneously constraining the flux density and the deconvolved size.  The free parameters are the peak intensity $I$ in Jy sr$^{-1}$, size of the emitting region, inclination $i$, position angle $PA$, and position offset.  The Markov Chain Monte Carlo (MCMC) Python package \textsc{emcee} \citep{emcee} is used to efficiently explore a wide range of parameters.  At each observed frequency, a first run is performed with 60 walkers ($10 \times$ free parameters) and 1000 steps.  Each walker explores the parameter space by gradually stepping into a region with the lowest $\chi^2$.  The best-fit values are obtained by calculating the mean of the last 100 steps of all walkers.  Then, a second run is performed with 120 walkers initialized by uniformly distributing them around the best-fit values obtained from the first run.  This second run is performed with 10000 steps to obtain the final set of best-fit parameters.  The best-fit values are determined statistically taking the mean of the last 1000 steps of the walkers with acceptance ratios greater than 20\%.  These best-fit values and their $1\sigma$ errors are listed in Table \ref{tbl:fitdust}.

The MCMC modelling of this high-fidelity data allow us to accurately determine that the TMC1A protostellar system is inclined at 50$^{\circ} \pm 3$ with a position angle of 75$^{\circ}\pm 4$.  The size of the dust disk is of the order of $0 \farcs 1$ ($FWHM = 0 \farcs 23$ or 30 au at 140 pc).  The deconvolved dust disk size determined from the high angular resolution ALMA data is smaller than the 100 au radius gaseous Keplerian disk \citep{harsono14}.  Most of the dust continuum flux density is emitted from the region probed by our ALMA data since the difference between the flux density obtained by ALMA and NOEMA data is small ($<20\%$) at 220 GHz.

\subsection{Molecular gas observations}
\label{sub:rottrans}

\begin{figure*}
 \centering
 \includegraphics[width=0.9\textwidth]{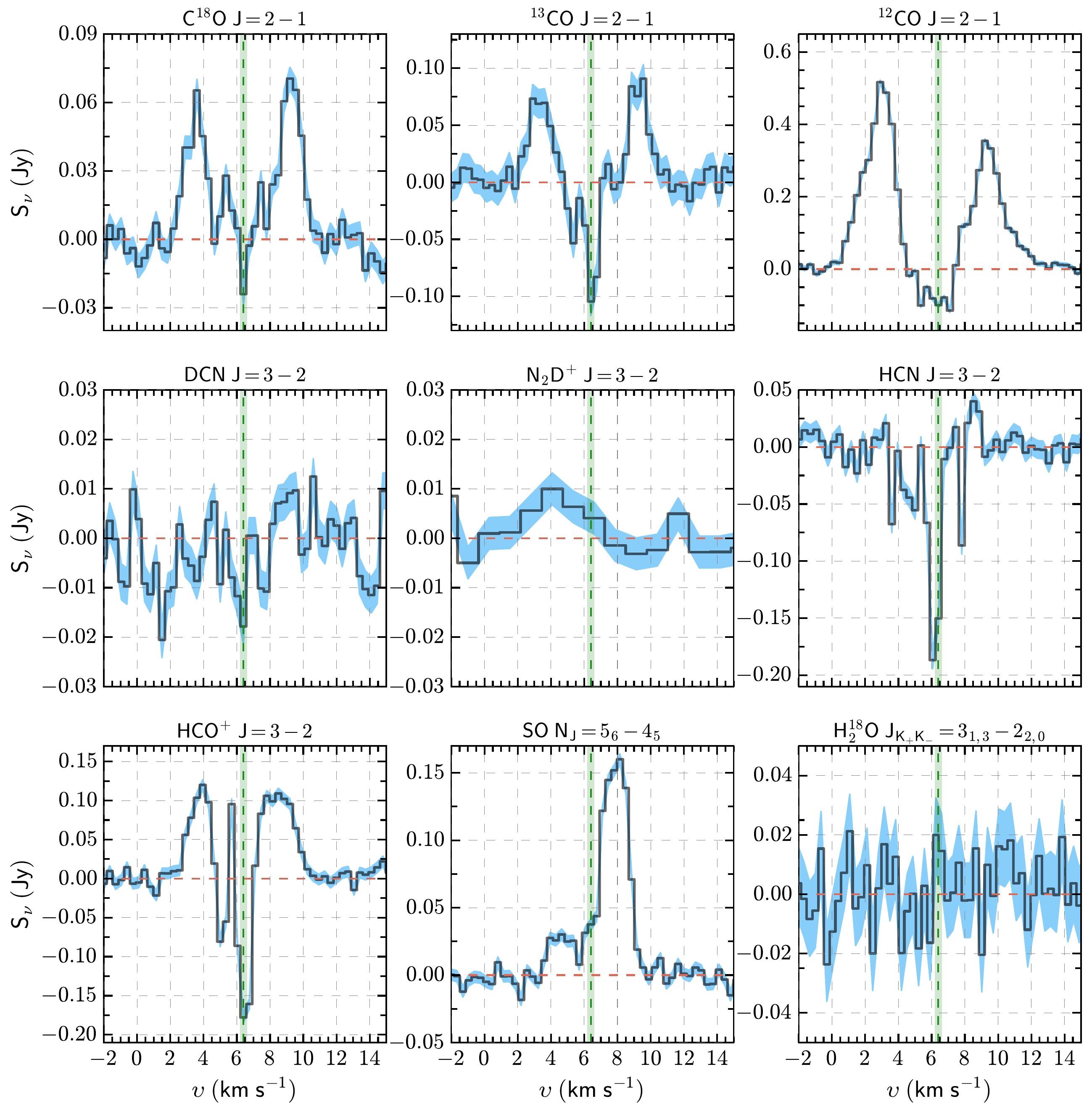}
 \caption{Observed molecular spectra from the inner 1$''$ square region centered on TMC1A.  The green vertical line is the systemic velocity of the system, while the horizontal, red dashed line indicates the baseline.  The 1$\sigma$ error is shown by the shaded blue regions. See Table~\ref{tbl:imgs} for more information on each line.
 }
 \label{fig:rotspectra}
\end{figure*}

\begin{figure*}
 \centering
 \includegraphics[width=0.95\textwidth]{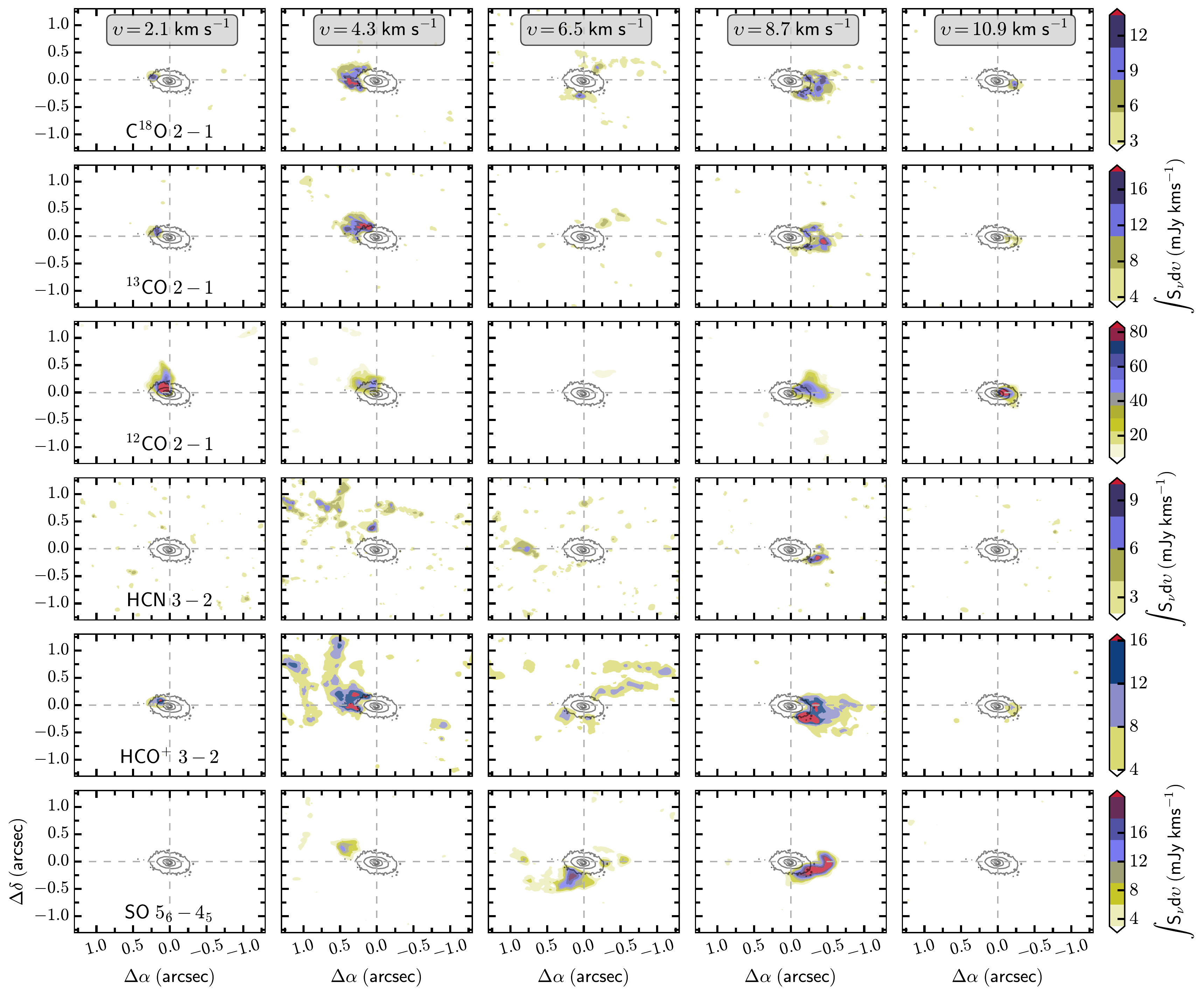}
 \caption{Channel maps of the six brightest molecular lines are shown:  \ceo\ $2-1$, \tco\ $2-1$, \mco\ $2-1$,  HCN $3-2$, HCO$^+$ $3-2$, and SO $5_6-4_5$ (top to bottom).  Each panel shows the line intensities integrated over six channels (left to right) and after clipping pixels with intensities $< 3\sigma$.  The grey box in the first row indicates the average velocity of the six channels. The color scale for molecular line is shown at the right hand side. The dust continuum emission of the aggregated observations is plotted linearly as black contours from $5\sigma$ to maximum intensity.  Beam sizes for each line can be found in Table~\ref{tbl:imgs}.
 }
 \label{fig:chanmaps}
\end{figure*}

\begin{figure*}
 \centering
\includegraphics[width=0.95\textwidth]{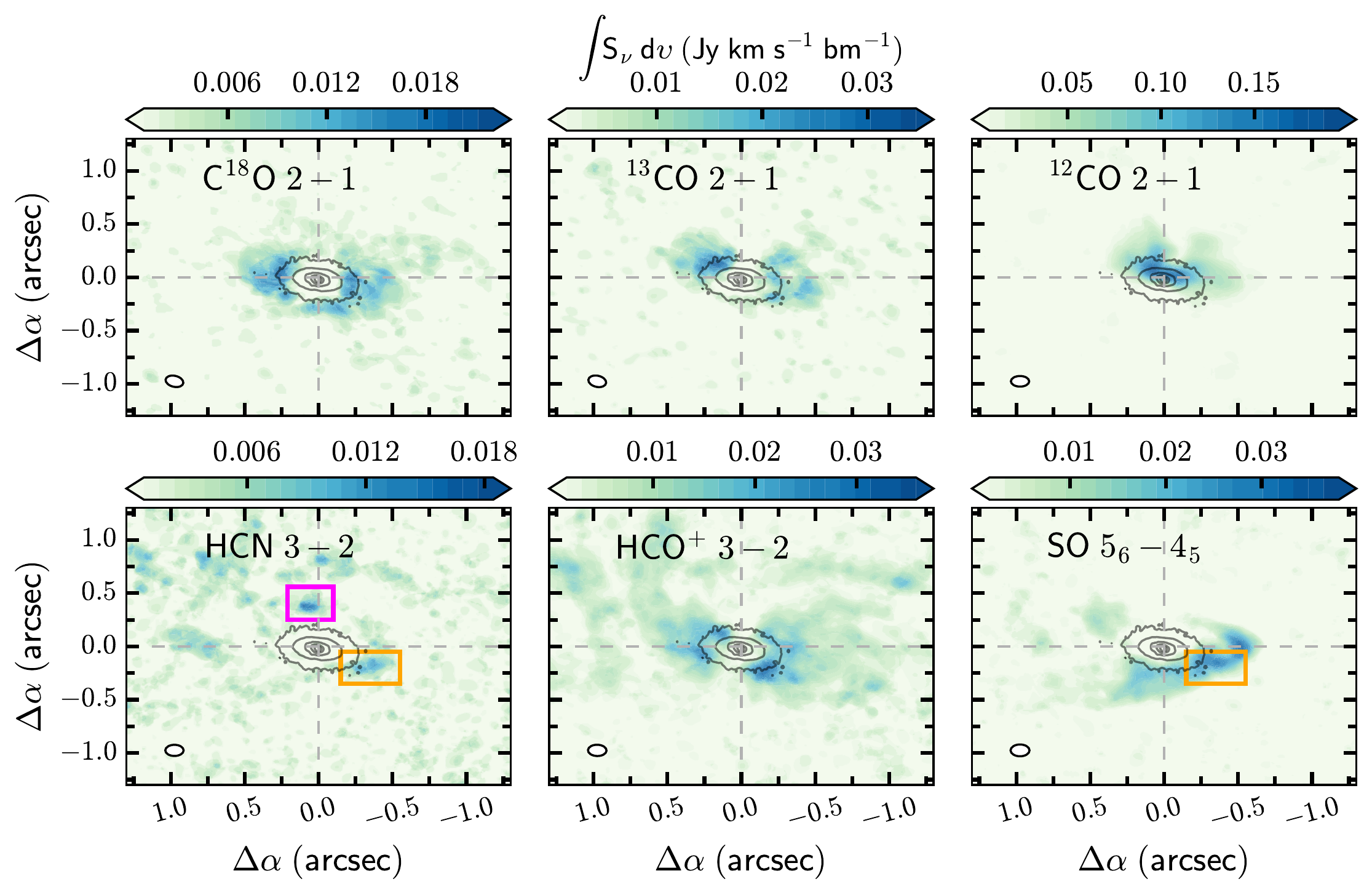}
 \caption{Zeroth moment maps of the six most strongly detected molecular lines.  They are calculated by integrating the emission from 1 to 12 \kms, considering only pixels where the emission $>3 \sigma$ (1$\sigma$ values are listed in Table~\ref{tbl:imgs}).  The synthesized beams are indicated in the lower left.  The dust continuum emission of the aggregated observations is plotted linearly as black contours from $5\sigma$ to maximum intensity.   There are two highlighted regions in the integrated HCN map, and one in the SO map: the supposed outflow component is shown in pink and the supposed disk component is in orange. See the text for details.
 }
 \label{fig:mom0maps}
\end{figure*}

In Fig.~\ref{fig:rotspectra}, we present spectra of the targeted molecular lines integrated over a 1$\arcsec$ box centered on TMC1A.  Out of the many targeted molecular lines (\mco\ $2-1$, \tco\ $2-1$, \ceo\ $2-1$, DCN $3-2$, \ce{N2D^{+}} $3-2$, HCN $3-2$, HCO$^{+}$ $3-2$, SO $5_6 - 4_5$, p-\heo\ 3$_{1,3}-2_{2,0}$, and Appendix ~\ref{app:undetected}), only a few are strongly detected in emission. While \ce{N2D^{+}} can be identified spectrally, its emission cannot be imaged with the current baselines coverage, suggesting that it is present in the diffuse large-scale envelope \citep[e.g.,][]{crapsi05a,tobin13}.  The channel maps and zeroth moment maps of the strong lines are shown in Figs.~\ref{fig:chanmaps} and \ref{fig:mom0maps}.

In the inner 1$\arcsec$ square aperture, \mco\ line emission is the strongest with a peak of 0.5 Jy. The peak line flux densities of \tco, \hcop, and SO are similar with a maximum of $\sim 0.1 $ Jy. The weaker lines are \ceo\ and HCN. The integrated flux densities are listed in Table~\ref{tbl:imgs} including the upper limits for a few molecular lines of interest.  The tabulated integrated flux densities are calculated considering pixels whose intensities $> 3 \sigma$ over the entire image.  The upper limits are calculated using a spatial mask ($x,y$) over the dust continuum emission following the methods outlined in \citet{carney19} assuming a Gaussian linewidth of 1 \kms\ (see \citealt{harsono20}).  We also report upper limits to the rotational transitions in the vibrational bending mode of HCN ($v_2 = 1$).  Their spectra are shown in Appendix~\ref{app:vibrational}.  The bending mode of HCN should be detectable toward the inner regions of protoplanetary disks in order to constrain the structure of the inner hot disk \citep[$T_{\rm gas} > 500$ K, ][]{bruderer15}.   The non-detection of these hot HCN lines limits the existence of hot gas to the inner 10 au of TMC1A.

The spectral line profile of \ceo\ $2-1$ is symmetric about the systemic velocity as would be expected from a rotating structure.  A similar line profile is seen for \tco\ $2-1$, while the \mco\ $2-1$ line shows a stronger blue-shifted emission reflecting the presence of the extended disk wind \citep{bjerkeli2016}.  The \hcop, SO, and HCN molecular lines exhibit narrow line profiles between $2 - 12$ \kms\ similar to \ceo\ and \tco.  The \ce{HCN} $J=3-2$ line has 6 hyperfine components \citep{mullins16} that could be responsible for the narrow absorption features (1-2 channels wide) seen in the spectrum at velocities significantly offset from systemic. Despite the weak HCN line profile, the channel maps in Fig.~\ref{fig:chanmaps} clearly indicate that its emission is detected in multiple channels.
%
%
\begin{figure}
 \centering
\includegraphics[width=0.45\textwidth]{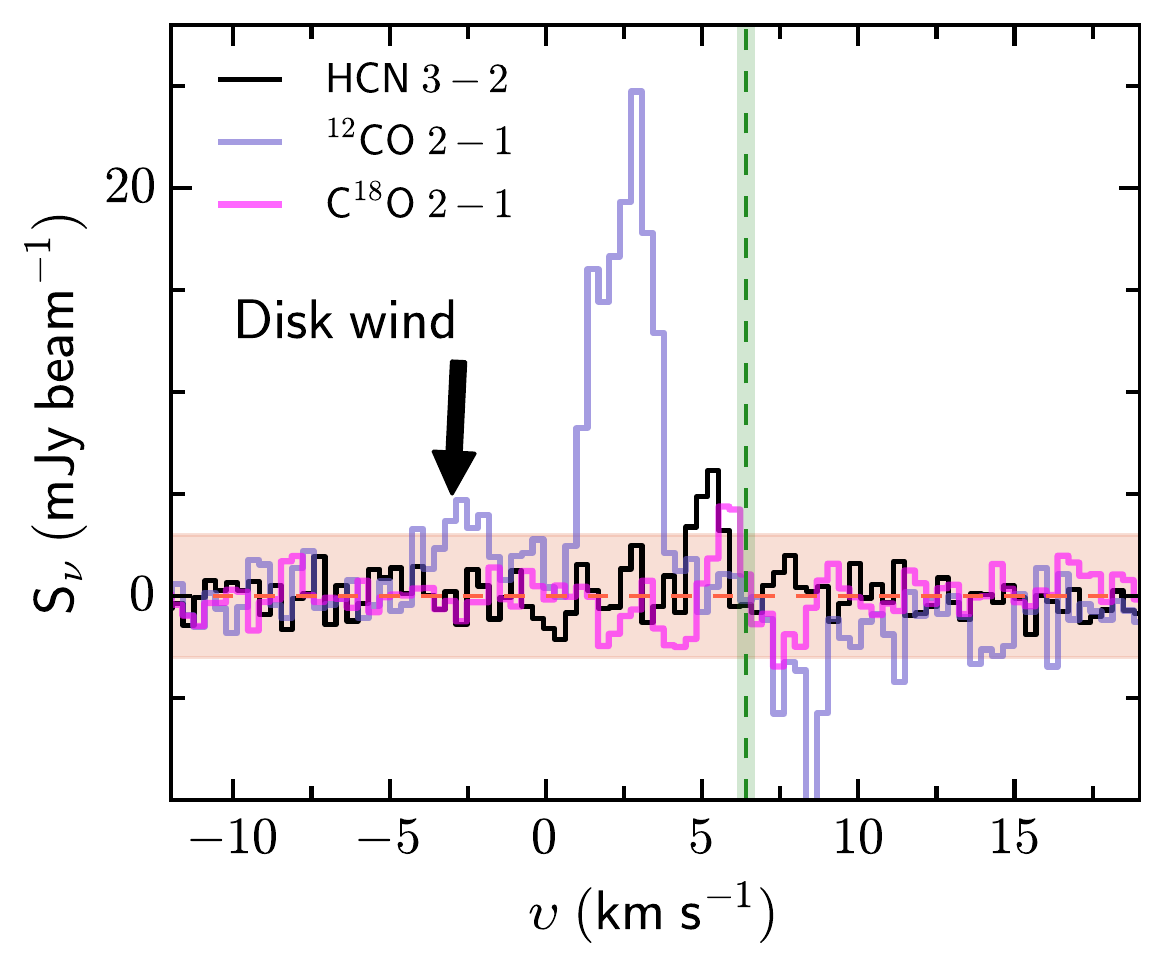}
 \caption{\ce{HCN}, \mco, and \ceo\ spectra from the region north of the disk, indicated by the pink rectangle in Fig.~\ref{fig:mom0maps}. The $-3$ to $3 \sigma$ levels for the HCN emission is indicated by the shaded region centered at 0 mJy beam$^{-1}$.  The outflow component seen in \mco\ is annotated.  The green vertical line denotes the systemic velocity.
 }
 \label{fig:hcnspectra43}
\end{figure}
%
%
\begin{figure}
 \centering
\includegraphics[width=0.45\textwidth]{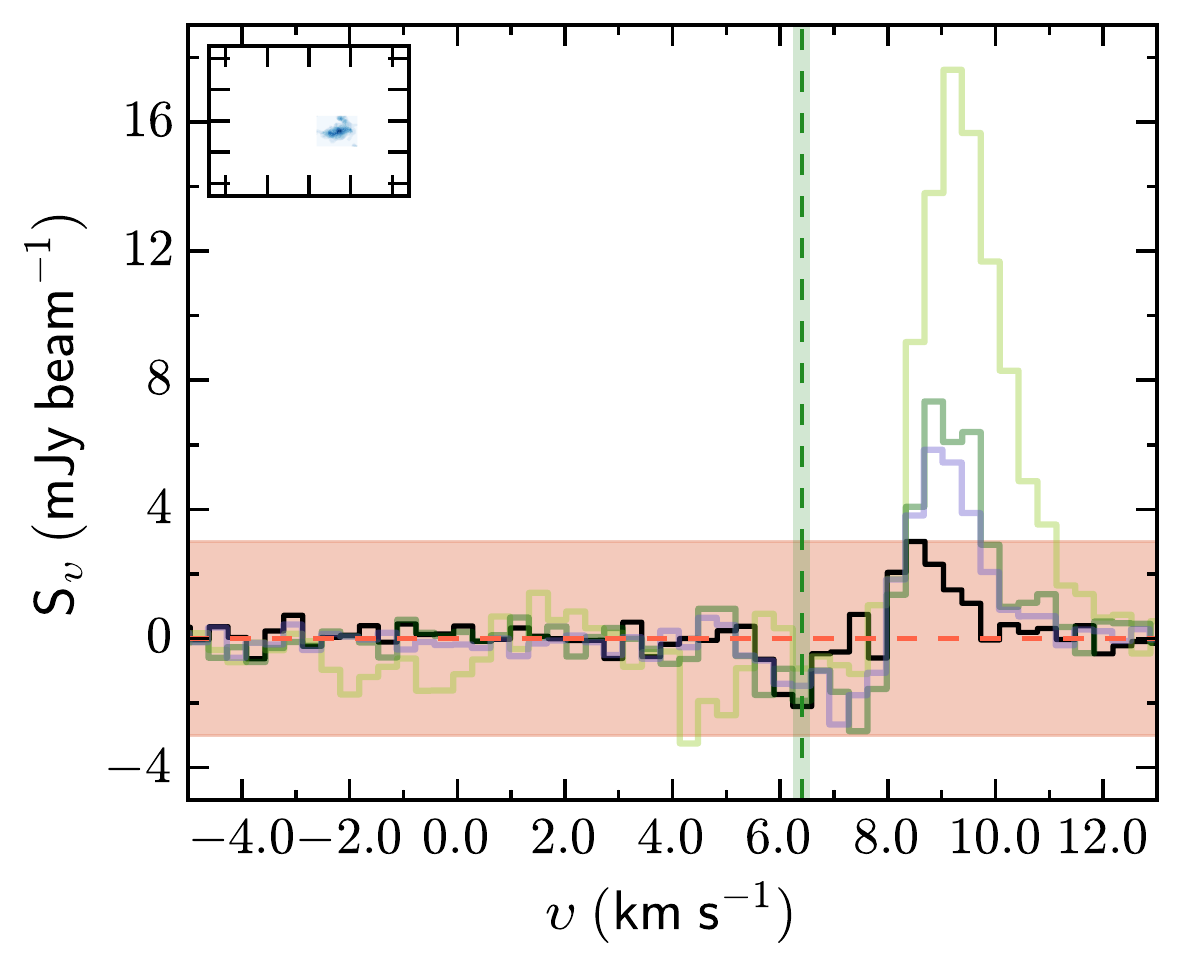}
 \caption{
 \ce{HCN}, \hcop, SO, and \ceo\ spectra from the western part of the disk (denoted by the orange rectangle in Fig.~\ref{fig:mom0maps}).  The shaded region highlights the $-3$ to $3 \sigma$ levels of the HCN emission, while the red horizontal dashed line shows the baseline, and the green vertical line shows the systemic velocity.
 }
 \label{fig:hcnspectra87}
\end{figure}

The channel maps in Fig.~\ref{fig:chanmaps} show that most of these molecular lines are detected in the vicinity of the millimeter dust continuum continuum emission.  \mco, \tco, \ceo, and \hcop\ show strong molecular emission from $1-12$ \kms\ while both HCN and SO are detected between $2.5 - 10$ \kms.  The \hcop\ and \ceo\ show emissions that are co-spatial.  The channel maps also show extended arc features in both \hcop\ and HCN lines that are due to filtered-out emission (Tychoniec et al. in prep.).  The zeroth moment maps (Fig.~\ref{fig:mom0maps}) show clearly these arcs.  The observed molecular lines avoid the central 30 au radius due to the optically thick dust and forms a ring-like structure \citep{harsono18}.  Meanwhile, the integrated SO emission is in a shape of a ring that extends up to $0\farcs5$ away, which has been observed toward other protostars \citep[e.g.,][]{sakai14nat}.

The observed HCN emission peaks at two locations that are marked by pink and orange boxes in Fig.~\ref{fig:mom0maps}. In order to investigate the origin of these HCN peaks, we extract an average spectrum over the two regions.  Figure~\ref{fig:hcnspectra43} shows the spectrum of HCN compared to \mco\ and \ceo\ in the region to the north of the disk (pink box).  The broad \mco\ emission indicates a molecular emission from the outflow cavity wall (with a peak at $\approx$2.5 \kms; \citealt{bjerkeli2016}) and a faster outflow/molecular wind component at $\approx$-3 \kms\ ($\approx v_{\rm sys} - 9$ \kms).  The narrow peak of HCN emission (at $\approx$5 \kms) is similar to that of \ceo.  The other region we highlight is located to the west (orange box).  Figure~\ref{fig:hcnspectra87} shows the comparison between HCN, \hcop, SO, and \ceo\ spectra in this region.  The similarity in their line profiles are indicating a common origin.

\begin{figure}
 \centering
\includegraphics[width=0.48\textwidth]{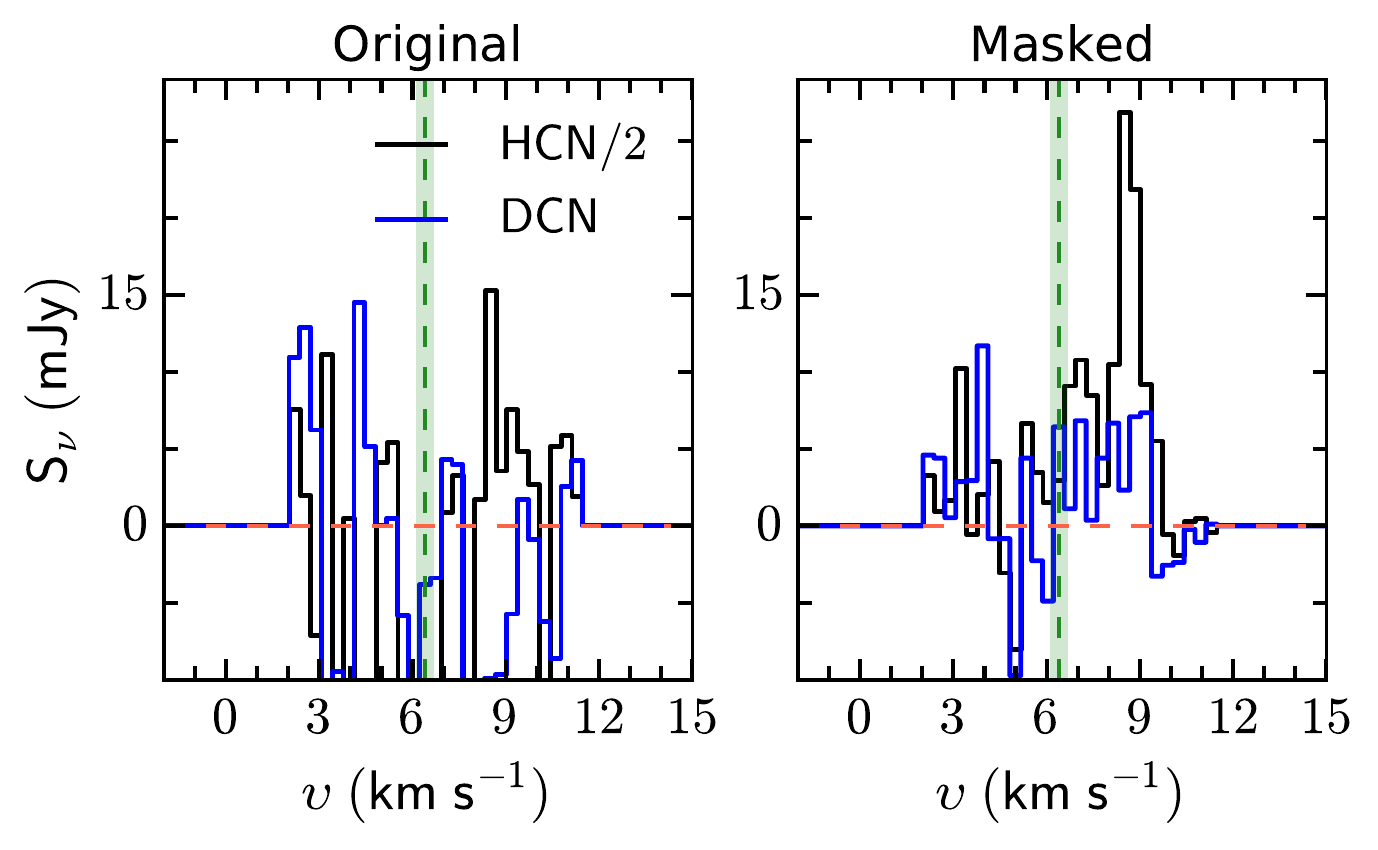}
 \caption{Comparison between the original HCN (black) and DCN (blue) spectra with the masked spectra.  The original spectra are the same as Fig.~\ref{fig:rotspectra}.  The mask is created from the \ceo\ channel maps (see text) to highlight the molecular emission associated with the Keplerian disk.
 }
 \label{fig:maskeddcn}
\end{figure}

In an attempt to extract the DCN emission, a proxy mask is created from the \ceo\ spectral cube by taking pixels ($x,y,v$) that are $>3 \sigma$.  Figure~\ref{fig:maskeddcn} demonstrates that it can extract the HCN emission corresponding to the red-shifted Keplerian disk at $v \sim 9$ \kms) as evidenced by its similarity to the \ceo\ spectra.  The DCN emission is, however, still not clearly detected over the same regions that \ceo\ is observed.

%
%
\section{Modeling results: molecular column densities and excitation conditions}
\label{sec:colden_excitation}

\subsection{Molecular column density of the disk}
\label{sub:diskcolden}

\begin{figure}
 \centering
 \includegraphics[width=0.5\textwidth]{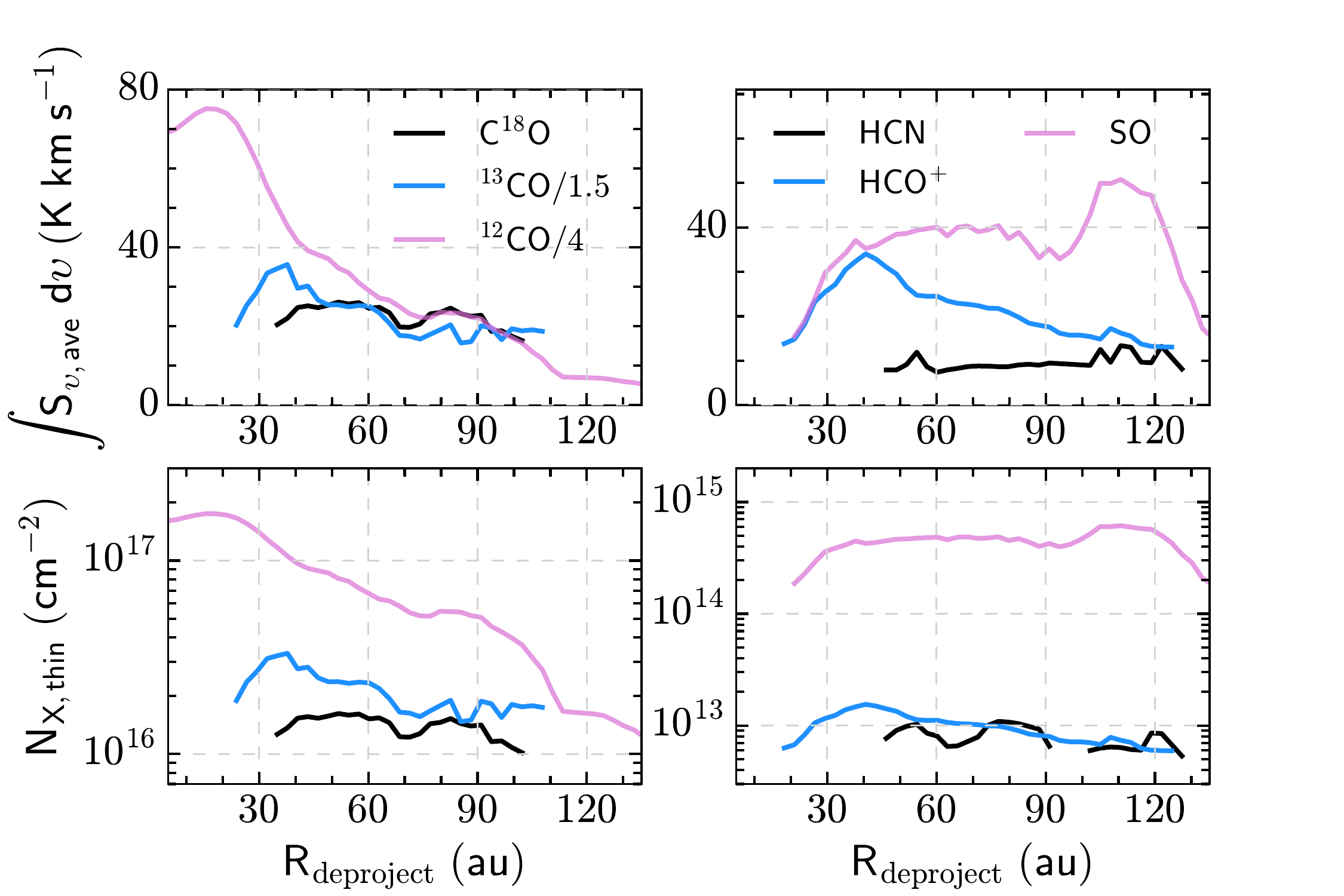}
 \caption{Azimuthally averaged, integrated line intensities (top) and column densities (bottom) as a function of the deprojected radius.  The column densities are derived assuming optically thin molecular line emission, and an excitation temperature of 30 K. The deprojection utilizes the $i$ and $PA$ obtained in Section~\ref{sub:dustcontinuum}.
 }
 \label{fig:columnave}
\end{figure}

The molecular abundance structure of the TMC1A disk can be derived from the spatially resolved molecular column densities.  To zeroth order, the molecular column density of the emitting gas can be determined by assuming a thermalized, optically thin layer of emission \citep{goldsmith99, mangum15}.   As an example, we first present the detailed analysis of the \ceo\ emission and compare the results from the optically thin approximation to a non-LTE radiative transfer analysis of a slab.  The same analysis is applied to the other molecular lines.

The integrated strength of the \ceo\ emission across the map ranges from 6 -- 15 mJy \kms\ per beam, which translates to 12 -- 30 K \kms.  In the optically thin approximation, adopting an excitation temperature of 30 K, the integrated column density of \ceo\ along the line of sight is then $\approx$ 10$^{16}$ cm$^{-2}$ ($\tau \approx 1$).  A temperature of 30 K is usually adopted to convert the dust thermal emission to dust mass and, in the case of \ceo\ at least, most of the molecular emission is emitted from a region where $T_{\rm gas} = T_{\rm dust}$ \citep[e.g.,][]{ceccarelli96, doty02}.  A non-LTE radiative transfer calculation of a slab using \textsc{RADEX} \citep{radex} is also performed as a check on the column density and excitation temperature.  These models predict a kinetic temperature of \ceo\ between 5--30 K with $\tau = 0.1 - 1$.  The $N_{\rm{C^{18}O}}$ is between $10^{15}$ to $10^{16}$ \ cm$^{-2}$. Therefore, the optically thin limit provides a reasonable estimate for the \ceo\ emitting layer.

Since the \ceo\ emission can be marginally optically thick ($\tau \approx 1$), line emission from the other observed CO isotopologs will definitely be optically thick ($\tau \approx 10$) simply from the isotopic ratios.  Therefore, we present the non-LTE calculations here, and adopt the CO collisional rate coefficients from \citet{corates}, obtained from the Leiden Atomic and Molecular Database (\textsc{LAMDA}; \citealt{lamda}).  These rates coefficients consider two collisional partners of ortho- and para-H$_2$ \citep{jankowski05}.  The integrated line intensity of \tco\ is between 16 -- 55 K \kms\ while it is 30 -- 300 K \kms\ for \mco.  The non-LTE radiative transfer model yields a suitable range of the excitation temperatures in between 15 to $\sim 70$ K to reproduce the observed integrated \tco\ emission with column densities between $10^{16}$ -- 10$^{17}$ cm$^{-2}$.  In the case of \mco\, the range of kinetic temperature is 50 to 300 K and \mco\ column densities between $10^{17}-10^{19}$ cm$^{-2}$ ($\tau = 1 - 7$).  Since both \tco\ and \mco\ lines are optically thick, the observed line emission comes from a surface layer (a slab) above the midplane.  This implies that the derived column densities are lower limits to the total column density of the gas that is present in the system.

The non-LTE radiative transfer calculations for HCN, \hcop, and SO lines also use the LAMDA collisional rate coefficients.  The rates for HCN from LAMDA are based on HCN-He collisions of \citet{hcnrate1}, scaled to H$_2$.  The collisional rates for \hcop\ are estimated using the dipole moment of \citet{botschwina93}.  SO-\ce{H2} collisional rates are scaled from SO-\ce{He} calculations of \citet{lique06}.  All of these rates consider a single collisional partner, H$_2$, without taking into account its ortho-to-para ratio.

In the optically thin limit, the observed HCN, \hcop, and SO emissions are produced by $N_{\ce{HCN}} = 5 \times 10^{12} - 10^{13}$ cm$^{-2}$, $N_{\ce{HCO^{+}}} = 5 \times 10^{12} - 10^{13}$ cm$^{-2}$, and $N_{\ce{SO}} = 10^{14} - 6 \times 10^{14}$ cm$^{-2}$, respectively, with an adopted excitation temperature of 30 K.  Figure~\ref{fig:columnave} shows the azimuthally averaged integrated line intensities and associated column densities in the optically thin limit (including \mco\ and \tco).  With the non-LTE slab model, the observed \hcop emission can be reproduced by kinetic temperatures between 30 -- 50 K and column densities of $10^{13}$ cm$^{-2}$, which are similar to the derived values presented in Fig.~\ref{fig:columnave} ($\tau \approx 1$).  Similar temperatures can also reproduce the observed HCN emission with $N_{\rm HCN} \approx 10^{12} - 10^{13}$ cm$^{-2}$, $\tau < 1$.  Finally, the observed SO emission can be reproduced with similar kinetic temperatures as the previous two molecules and higher column densities of $10^{14} - 10^{15}$ cm$^{-2}$.  The physical conditions of SO along the line of sight are consistent with \tco, which is optically thick.  The optically thin local thermodynamic equilibrium (LTE) calculations can provide useful limits to the the column densities of the detected molecules.

Since we have observed the same rotational transition in HCN and DCN, albeit only an upper limit on the DCN emission, we can place constraints on the D/H ratio in the vicinity of TMC1A. We restrict this analysis to the disk-averaged value by determining the column density and temperature of the upper limit to the integrated line intensity obtained from the HCN and DCN spectra, summed over the Keplerian disk.  As a check, we determined that the physical conditions derived from the averaged HCN spectrum are similar to the azimuthally-averaged values in the previous section.  For a range of excitation temperatures between 15 to 50 K, an average value of $N_{\ce{HCN}} \sim 1.1 \times 10^{12}$ cm$^{-2}$ is derived for the TMC1A disk.  The same analysis provides an upper limit for the DCN column density of $\sim 3 \times 10^{11}$ cm$^{-2}$, resulting in a DCN$/$HCN ratio of $<$0.3.  If we instead only consider the HCN emission between 8 to 10 \kms\ (where the HCN emission is strongest), the upper limit of DCN$/$HCN is then $<0.26$.

\subsection{Temperature structure of the disk}
\label{sub:tempstructure}

\begin{figure}
 \centering
 \includegraphics[width=0.48\textwidth]{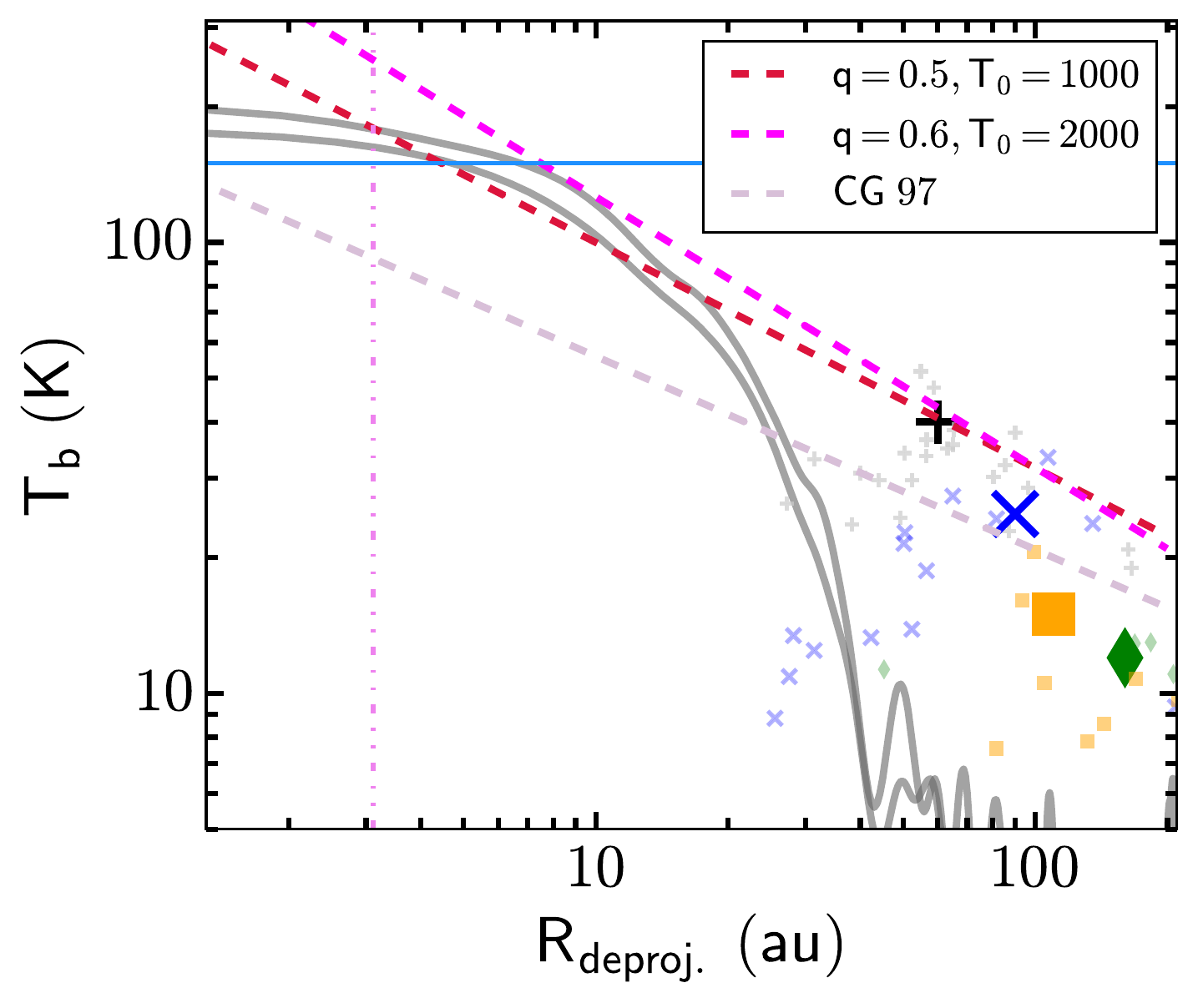}
 \caption{Brightness temperature profile along the disk major axis as a function of deprojected radius.  Gray solid lines show the observed dust continuum brightness profile at 230 and 240 GHz.  Peak brightness temperatures of \tco\ (black $+$), \hcop (blue crosses), \ce{HCN} (green diamonds), and \ce{SO} (orange squares) are also plotted.  The smaller symbols show the observed peak temperatures at velocities between 1-4 \kms\ and 9-12 \kms\ that contain $>3 \sigma$ emission, while the large symbols indicate the averaged values of these points.  For comparison, a power-law ($T = T_0 \left ( R/0.1 \ \mathrm{au} \right )^{q}$) and the \citet[][; CG97]{chiang97} temperature profiles are also shown (dashed lines).  The values of $T_0$ and $q$ are indicated in the legend.  The vertical purple dashed-dot line indicates $0.5 \times$ the beam size, while the horizontal blue line denotes the water snowline at 160 K \citep{meijerink09}.
 }
 \label{fig:TMC1Atemp}
\end{figure}

Spatially resolved observations of dust continuum and molecular line emission can be used to estimate the temperature structure of the disk.  To this end, Figure~\ref{fig:TMC1Atemp} shows the observed dust continuum brightness temperature at 230 and 240 GHz, as well as the molecular line peak brightness temperatures, as a function of the deprojected radius.  The flattening in the inner $<10$ au is caused by unresolved continuum emission. If the dust emission is optically thick within the inner 30 au, however, the peak continuum brightness temperature gives indications on the disk's temperature profile.

Using the high-velocity channels ($1 < v < 4$ \kms\ and $9< v <12$ \kms; to avoid resolved-out emission near the systemic velocity), the peak brightness temperature of \tco, \hcop, HCN and SO molecular lines as a function of the deprojected radius is shown in Fig.~\ref{fig:TMC1Atemp}.  There are only small differences between the \tco\ and \ceo\ brightness temperatures, and only at $> 30$ au radius, so only \tco\ is plotted.  After acknowledging the decrease in molecular emission in the inner 30 au radius, as well as optically thin components, the remaining few optically thick emission data points can provide the additional constraints on the disk temperature profile.  Their approximate average values and locations are indicated by the large symbols in Fig.~\ref{fig:TMC1Atemp}.

The equilibrium midplane temperature of a circumstellar disk can be characterized using a power-law in radius with an index between -0.4 to -0.5 \citep{chiang97,vanthoff18b}.  We find that, by eye, a temperature profile given by $2000 \times \left ( R/0.1 \ {\rm au} \right )^{-0.6}$ seems to be consistent with our dust and gas observations of TMC1A.  It reproduces the dust brightness temperature profile in the inner 30 au and intersects the peak of the \tco\ emission.  The derived temperature structure is similar to the disk around L1527 ($L_{\rm bol} = 1.9-2.6 \ L_{\odot}$) as measured by \citet{vanthoff18b}.   Furthermore, the slope of the temperature profile is steeper than expected from purely re-radiated stellar photons ($q\simeq -0.4$), implying that the observed emitting surface shifts from the hot disk surface to the cold disk/envelope gas at large radii.  The estimated temperature of 1000-2000 K at 0.1 au is also consistent with the observed $L_{\rm bol}$ ($2.7 \ L_{\odot}$) for a 4000 K protostar ($R_{\star} \sim 3.4 $ $R_{\odot}$).  The derived temperature structure implies that the warm TMC1A disk does not have a CO freeze-out region ($T_{\rm dust} = 30$ K) in the inner 100 au of the disk \citep{vanthoff18b}.  Recent \ce{C^{17}O} observations presented by \citet{vanthoff20} also indicate the lack of CO freeze-out region in TMC1A disk.

%
%
\section{Discussion: the molecular layer of disks}
\label{sec:molecularlayer}

The unprecedented spatial resolution of these observations allows us to disentangle molecular emission from the disk and from the large-scale envelope.  The simple molecules targeted in our data set are \mco, \tco, \ceo, SO, \ce{HCO^{+}}, HCN, and DCN, plus the spectrally unresolved \ce{N2D^{+}}.  These molecules can be used to better understand the evolution of the physical and chemical structure of disks during the embedded stage of star and planet formation.

\subsection{Physical components of embedded protostars traced by molecular lines}
\label{sub:physicalcomponents}

\begin{figure*}
 \centering
\includegraphics[width=0.98\textwidth]{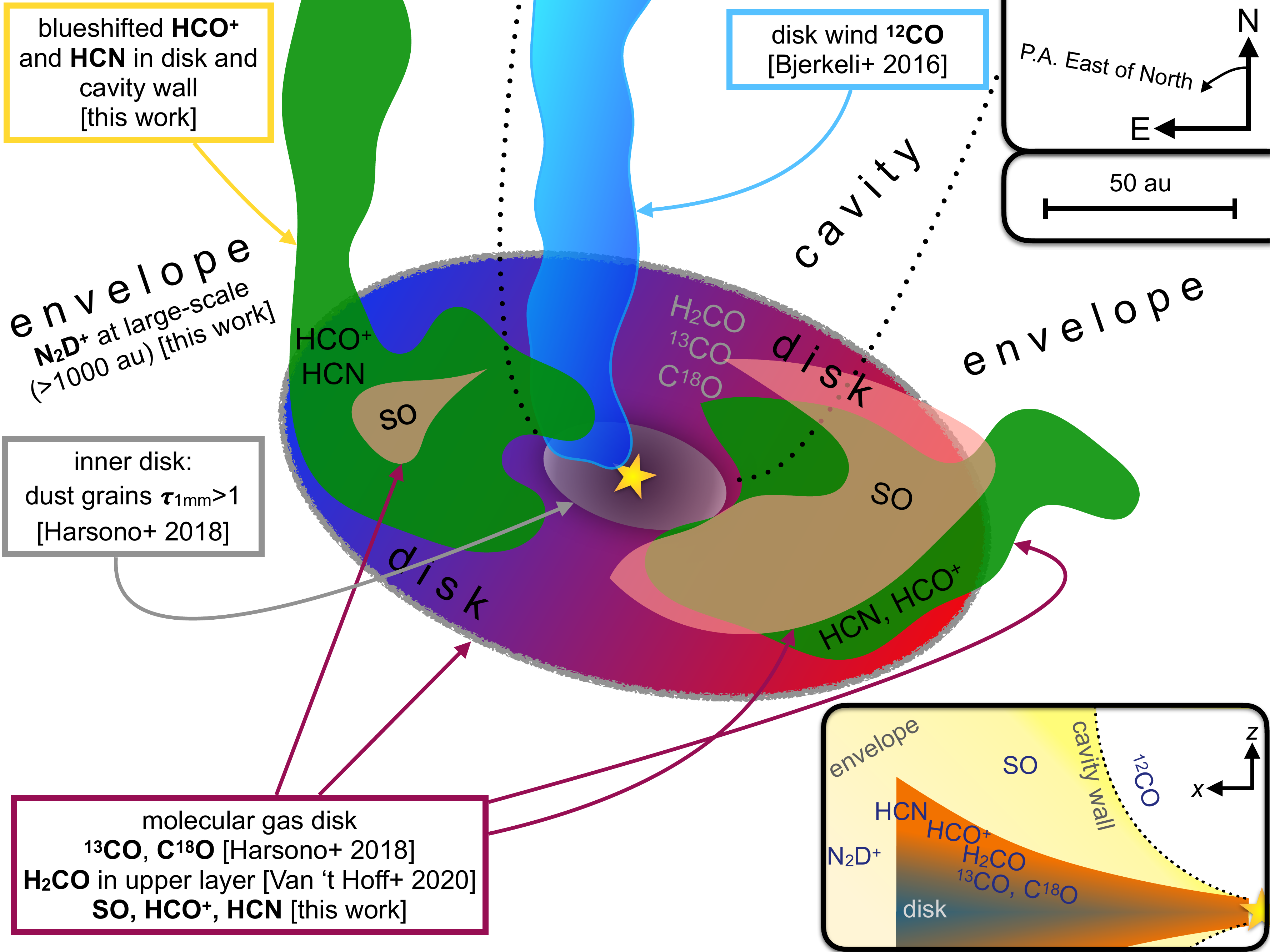}
 \caption{A schematic of the observed molecular emission and the physical components of the embedded protostellar system TMC1A.  For the references given in the illustration, `et al.' is abbreviated as `+'. References: \citet{bjerkeli2016}, \citet{harsono18}, \citet{vanthoff20}. Main panel: molecular gas components projected in the same way as in the observation. Inset at bottom right: interpretation of relative location of the origin of molecular gas emission shown in the ($x$,$z$) plane, with $z$ along the rotation axis of the disk.
 }
 \label{fig:cartoon}
 \end{figure*}

The observed molecular lines trace different physical components of an embedded protostellar system, i.e., the inner envelope ($R < 500$ au), Keplerian disk, disk wind, and the cold, infalling envelope.  The schematic in Fig.~\ref{fig:cartoon} summarizes the observed molecular tracers, their emitting regions, and the physical component that we associate each tracer to.  As done previously, we use the symmetric line profile of \tco\ and \ceo\ 2--1 to establish the location of the Keplerian disk.  We then compare the other molecular line profiles with \tco\ and \ceo\ to provide a first look into the physical components traced by these other lines in TMC1A.  Finally, we add the channel maps and zeroth moment maps comparison to give the complete understanding on the emitting regions of these molecules and the structure of the TMC1A system.

From the line profiles in Fig.~\ref{fig:rotspectra}, the roughly symmetric \hcop\ 3--2 line profile indicates that it is tracing the Keplerian disk.  This is not unexpected since \hcop\ is mainly formed through reactions that involve CO (see Section~\ref{sub:accretion}), and, thus, \hcop\ emission should coincide with the CO emission.  On the surface of the disk with low densities, chemical reactions tend to favor the production of \ce{N2H^{+}} \citep{aikawa15} and the destruction of \hcop\ by water vapor \citep{jorgensen13,vanthoff18a}.  The emitting \hcop\ layer in the disk is confined to a small outer layer (see \citealt{mathews13}).

The asymmetric line profiles of HCN and SO, meanwhile, strongly indicate that these lines are tracing multiple physical components along the line of sight (Fig.~\ref{fig:hcnspectra87}). For example, both HCN and SO show red-shifted emission to the west that corresponds (both spatially and in velocity space) to the Keplerian disk as traced by \ceo\ and \hcop.

The narrow line profile of HCN 3--2 (Figs.~\ref{fig:hcnspectra43} and \ref{fig:hcnspectra87}) strongly indicates emission from quiescent gas, and this gas could be in either the envelope or Keplerian disk.  The similarity in the line shape and velocity peak of \hcop\ and \ceo\ in Fig.~\ref{fig:hcnspectra87} points to the Keplerian disk as the origin of the red-shifted HCN emission. At lower velocities, the HCN line profiles shows evidence of some contribution from the surrounding envelope.  Therefore, we conclude that the HCN line traces the disk-envelope transition region.  A detailed 2D physical and chemical model is required to quantify the exact location of the molecular emission, but this is beyond the scope of the current work.  Finally, the morphology and velocity of \ce{HCN} 3--2 emission (Figs.~\ref{fig:mom0maps}, \ref{fig:hcnspectra43}) does not appear to be connected to the outflow observed previously in CO \citep{bjerkeli2016}.

\begin{figure}
 \centering
 \includegraphics[width=0.5\textwidth]{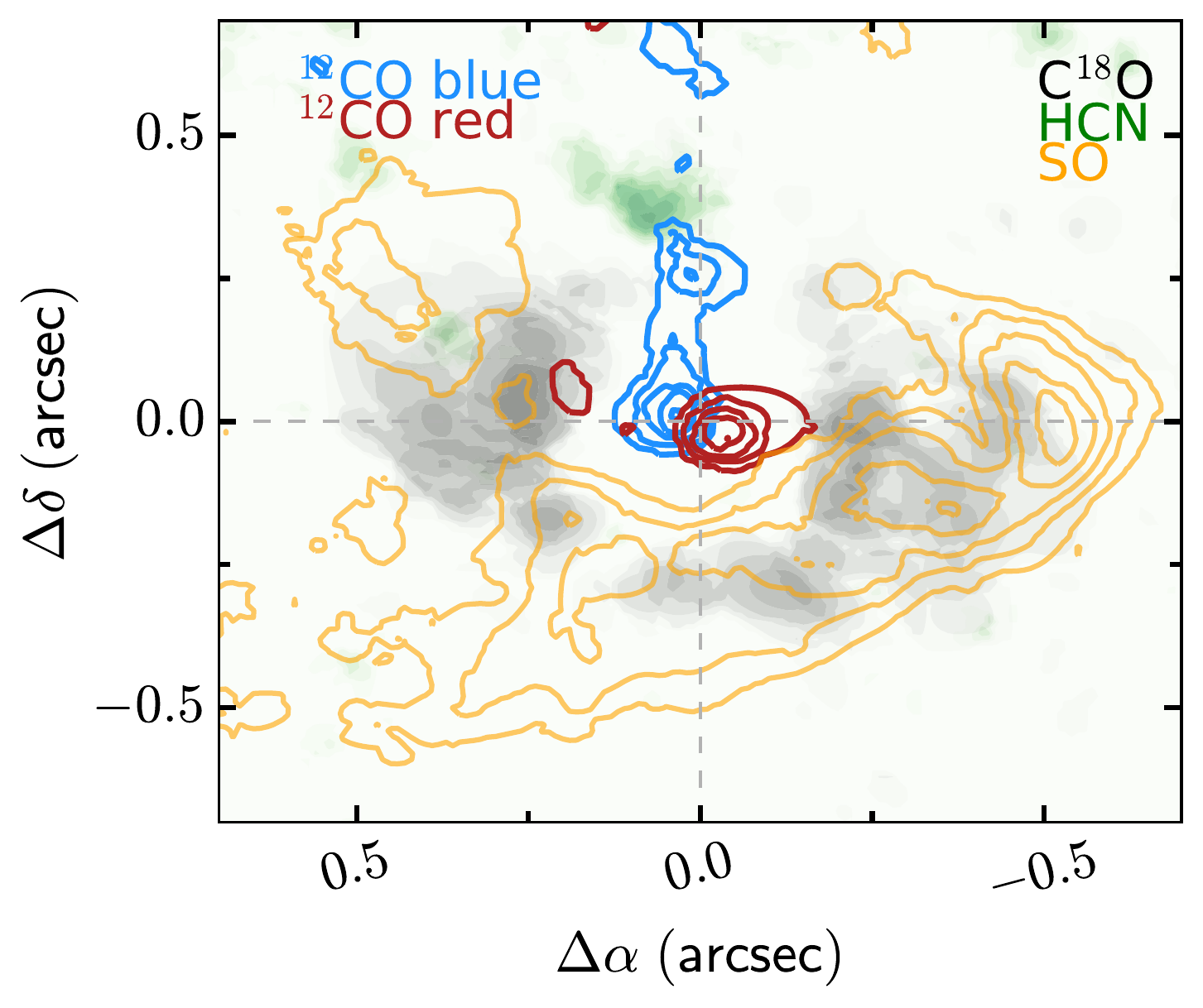}
 \caption{Comparison of SO, \ceo, \ce{HCN}, and \mco\ emitting regions using zeroth moment maps. The \ceo\ emission is shown in grayscale with a linear scaling between 0.2 to 0.8 of the maximum in order to highlight the disk component.  Green filled contours shows HCN integrated from -1 to 6 km s$^{-1}$. The color spans linearly from 0.3 to 1.0 of the maximum value.  Blue-shifted \mco\ (-7 to -1 km s$^{-1}$) and red-shifted (13 to 19 km s$^{-1}$) emission are highlighted by the blue and red contours, respectively.  The contours span linearly from 0.15 to 1.0 of the maximum.  Finally, the SO integrated emission is shown using orange contours with a linear scaling from 0.15 to 1.0 of the maximum.
 }
 \label{fig:TMC1ASO}
\end{figure}

Our spatially resolved SO observations show morphological features (Figs.~\ref{fig:chanmaps} and \ref{fig:mom0maps}) that are similar to CO and \hcop.   The narrow line profile of SO in Fig.~\ref{fig:hcnspectra87} and the low peak temperatures of SO in Fig.~\ref{fig:TMC1Atemp} meanwhile rule out an accretion shock origin \citep[e.g.,][]{sakai14nat}.  A comparison between the zeroth moment maps of \ceo, HCN, and \mco\ is shown in Fig.~\ref{fig:TMC1ASO}.  Since the SO and \mco\ molecular emission are not co-spatial, it excludes a disk wind origin for SO \citep[e.g.][]{tabone17} for TMC1A.  However, the blue-shifted HCN emission in Fig.~\ref{fig:TMC1ASO} peaks at the location near the blue-shifted CO emission suggesting a region where the wind and the envelope interact.  The combination of the SO line profile and its peak brightness temperature (Fig.~\ref{fig:TMC1Atemp}) indicates that it originates from the infalling, warm inner envelope along the line of sight.  The location of the SO emission with respect to the disk wind (blue-shifted emission at 4.3 \kms) and the Keplerian disk further supports that SO is located along the outflow cavity wall.  This is also hinted by the presence of lower velocity component in the spectrum shown in Fig.~\ref{fig:hcnspectra87}.  The favorable orientation of TMC1A and these deep spectrally resolved molecular line observations allow us to disentangle the different physical components of this embedded disk. Such an empirical derivation of the origin of SO would have been impossible in an edge-on system and without additional gas line tracers.


\subsection{Molecular abundances in the TMC1A disk.}
\label{sub:molabunds}

\begin{figure*}
 \centering
 \includegraphics[width=0.95\textwidth]{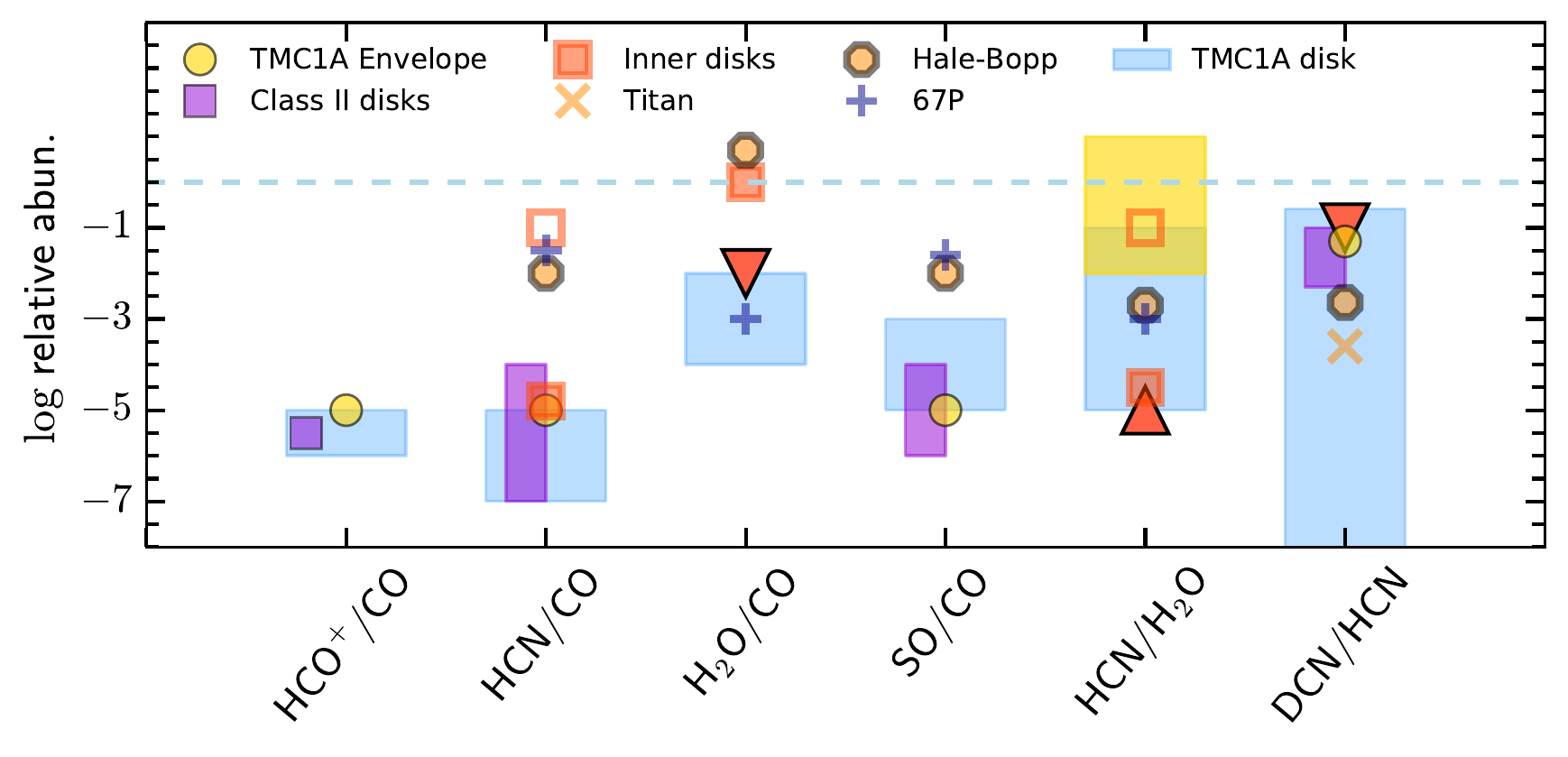}
 \caption{Comparison of the relative molecular gas abundances in the TMC1A disk and its envelope, Class II disks, and Solar System bodies.  The range of values obtained in this work is shown by the blue shaded rectangles.  The red arrows indicate lower and upper limits.  Yellow circles indicate the molecular abundance ratios of the protostellar envelope around TMC1A from \citet{jorgensen04c}.  The abundances of HCN and DCN in the envelope are approximated to the values in \citet{schoier02}, \citet{roberts02}, and \citet{tafalla06}.  The water gas fraction in the protostellar envelope is assumed to be between $10^{-7}$--$10^{-4}$ based on models by \citet{caselli12b} and \citet{schmalzl14}.  Molecular abundances in Class II disks (purple rectangles) are based on values from \citet{mathews13}, \citet{teague15}, \citet{guzman15}, \citet{cleeves18}, \citet{huang17}, and \citet{booth19}.  The inner disk hot gas molecular abundances (empty red squares) are adopted from \citet{najita13} and \citet{salyk11} shown by filled red squares.  The abundances of comet Hale-Bopp are based on \citet{bockelee00}.  The abundances of comet 67P/Churyumov?Gerasimenko are adopted from \citet{rubin19}.  The DCN/HCN fraction for Titan and comet Hale-Bopp are obtained from \citet{molter16} and \citet{meier98}, respectively.
 }
 \label{fig:TMC1Amolabun}
\end{figure*}

One of the major questions in star and planet formation is whether or not planet-forming materials enter the disk from the envelope unaltered.  Alteration of planet-forming materials require high temperatures that can generally only be achieved in interstellar shocks and the inner regions of protostellar systems.  These changes would be relevant for the accretion history of the protostar (e.g., dust sublimation in the inner regions of the disk or protostellar accretion bursts) and disk formation (e.g., accretion shocks at the disk-envelope interface that affect the chemistry and ionization fraction) that can be explored by comparing the chemical abundances of young stellar objects at different scales.  Indeed, differences in chemical content between the embedded disk and its surrounding envelope may point to on-going physical processes during the early stages of star and planet formation.

The molecular line spectra and the kinetic temperature analysis presented in this work identify some regions and velocity intervals that emission from the disk, but also indicate that some lines of sight toward TMC1A are contaminated by its surrounding envelope.  We showed in Sect.~\ref{sub:diskcolden} that assuming the optically thin limit can provide useful constraints on the emitting molecular layer and the column densities of the disk.  Furthermore, with the azimuthally averaged column densities in hand, we can constrain the molecular abundances in the disk.  In addition, since these observations use the longest baselines configuration of ALMA, they are only sensitive up to a maximum recoverable scale of $\sim 0\farcs5$ ($\sim$ 70 au).  Thus, the observed molecular emission is restricted to the inner hundred au of TMC1A, however, the chemical timescales in these regions are indeed longer than the infall timescale (the time for the gas to fall onto the disk).  Therefore, we expect that the derived molecular abundances on the observed scales of these observations should not significantly change before the molecular gas enters the disk.

By comparing the derived column densities of the different species, we obtain $N_{\rm HCO^{+}}/N_{\rm ^{13}CO} \sim 10^{-4} - 10^{-3}$ and $N_{\rm SO}/N_{\rm ^{13}CO} = 10^{-3} - 10^{-1}$.  The HCN abundance is meanwhile estimated through comparison with the \ceo\ column density, giving $N_{\ce{HCN}}/N_{\ce{C^{18}O}} = 10^{-4}-10^{-2}$. The proxy mask (see Sect.~\ref{sub:rottrans}) using the \ceo\ spectral cube recovers more HCN flux than a proxy mask using the \tco\ spectral cube.  Although some envelope material is contaminating the detected emission, it is still useful to calculate the relative abundances for the disk, which are shown graphically in Fig.~\ref{fig:TMC1Amolabun}.  Adopting the ISM isotopic ratios of $\ce{^{12}C}:\ce{^{13}C} = 70$ and $\ce{^{16}O}:\ce{^{18}O} = 540$ \citep{wilson94}, the derived molecular abundances relative to \mco\ are then $X_{\ce{HCO^{+}}}/X_{\ce{CO}} \sim 10^{-6}-10^{-5}$, $X_{\ce{HCN}}/X_{\ce{CO}} \sim 10^{-7}-10^{-5}$, and $X_{\ce{SO}}/X_{\ce{CO}} \sim 10^{-5}-10^{-3}$.

It has been demonstrated that the CO gas abundance in young disks remains close to the canonical ISM value of $X_{\ce{CO}} =$ $10^{-4}$ relative to \ce{H2} \citep{harsono14, vanthoff18b, kzhang20}.  Adopting this value, we estimate abundances of $ X_{\ce{HCO^{+}}} = 10^{-10}$--$10^{-9}$, $X_{\ce{HCN}} = 10^{-11}$--$10^{-9}$ and $X_{\ce{SO}} = 10^{-9}$--$10^{-7}$ in the disk of TMC1A, where $X$ denotes an abundance with respect to \ce{H2}.  One should  keep in mind, however, that these abundance fractions are valid in the emitting regions of the inner warm envelope and the layers of the young disk above the dust photosphere.

Remarkably, the gas abundance ratios in the disk around TMC1A are similar to its protostellar envelope (Fig.~\ref{fig:TMC1Amolabun}).  This implies a smooth transport of materials without the presence of strong shocks ($T > 100$ K) that modify the molecular composition of the material en route to the disk \citep{visser09}.  \hcop\ and \ce{HCN} show features of filtered out molecular emission.  The comparison of the total \ceo\ and \tco\ integrated emission between this study and \citet{harsono14} also indicate 90\% of the emission is filtered out as a result of going from $0\farcs5$ to $0\farcs1$ resolution.  However, the lack of absorption in the SO line profile suggests that our studies recover most of its emission.  Thus, the range of the SO/CO abundance ratio shown in Fig.~\ref{fig:TMC1Amolabun} may instead be explained by filtered out CO emission.   It further indicates that \hcop and SO are not drastically affected by the physical processes that take place during disk formation.

The chemical composition of the TMC1A disk is also similar to that of Class II disks.  The main difference is the lower absolute CO gas abundance that is inferred from CO observations of Class II disks \citep[e.g.,][]{miotello17,kzhang20}.  CO depletion during star and planet formation is linked to the formation of complex organics, on-going planet formation, and the UV field around low-mass stars \citep[e.g.,][]{moyu17,bosman18,dodson18,mcclure19}.  As such, gas abundance ratios (as compared to absolute abundances) may be less sensitive to variations between objects and better reflect how chemistry changes with evolutionary state.  Therefore, the similar gas molecular abundance ratios between the TMC1A disk and Class II disks may indicate that the molecular layer of protoplanetary disks do not change significantly once the disk is formed ($M_{\rm d}/M_{\rm env} \sim 0.5$).

Interestingly, for the most part, the abundance ratios between protostellar systems and Solar System objects do show significant differences.   In order to assess whether SO is truly being affected by disk formation, deep observations of other S-bearing species (e.g., \ce{H2S}, \ce{SO2}, \ce{CS}) in combination with multi-resolution CO observations are needed to recover the total emission from the inner 500 au of the TMC1A system.  Without observations of other S-bearing molecules, it is difficult to conclude the origin of this discrepancy.  A comparison between S-bearing species in IRAS 16293--2422 and comet 67P imply that the Solar System formed in a higher UV field than IRAS 16293--2422 \citep{drozdovskaya18} since S is a volatile element and sensitive to the changes in the UV field.  Thus, it would seem that the differences in S-bearing molecules may trace the strength of the UV field during protostellar evolution.

With regards to HCN and DCN, \citet{huang17} show that there is a spread in DCN abundances in disks that stems from the different cold and warm routes of DCN formation \citep[see][and references therein]{aikawa99,willacy07}.  Unfortunately, neither HCN nor DCN were detected toward TMC1A in the single-dish survey of \citet{jorgensen04c} which hinders us from a straight comparison between disk and envelope values; the ratio in Fig.~\ref{fig:TMC1Amolabun} instead adopts the prestellar core value.  Our upper limits to the DCN/HCN fraction in the TMC1A disk are, meanwhile, consistent with other various young stellar objects and Solar System objects.  A small variation in DCN/HCN in protostellar envelopes has been observed \citep{jorgensen04c}, which seems to be related to the bolometric luminosity of the source.  However, it is still not well understood whether the level of deuteration is modified during disk formation.  For example, an older disk presented in \citet{huang17} seems to have a DCN/HCN ratio consistent with comet Hale-Bopp, which favors the idea of an evolving DCN/HCN ratio. Meanwhile, the DCN/HCN ratio of Titan is different than that of both disks and comets, indicating a different origin for the observed deuteration fraction.

A large variation across different types of objects is also seen in \ce{HCN} and \ce{H2O} abundance ratios.  Interestingly, abundance variations in both \ce{H2O} and HCN have been linked to angular momentum transport \citep[e.g,][]{cuzzi06a,kalyaan19} and planet formation \citep[e.g.,][]{najita13, krijt16, du17, najita18}.  Our sub-mm \ce{H2O} column density for TMC1A is adopted from the upper limit in \citet{harsono20} ($N$(H$_2$O)/$N$(H$_2$) $< 10^{-6}$) that probes the inner 10 au of the disk.   Hot water observations in the mid-IR around Class II disks \citep[e.g.,][]{carr08, salyk11,najita13} meanwhile probe the inner 2 au where terrestrial planets are expected to form.  The observed correlation between the \ce{HCN}/\ce{H2O} mid-IR line flux ratios and the disk mass (via the millimeter flux density) has been suggested to trace planetesimal formation inside of 5 au \citep{najita13, najita18}.  If the observed mid-IR line flux ratios correspond to the relative abundances in the emitting layers \citep[$z/R \sim 0.6$, ][]{bruderer15}, then they are consistent with the gas molecular abundance ratios derived in this work.  The differences between the abundance ratios in the large-envelope and disk (Fig.~\ref{fig:TMC1Amolabun}) thus suggests that \ce{HCN}/\ce{H2O} is set during disk formation, which is indeed supported by the similar abundance ratio observed in comets 67P and Hale-Bopp.  On the other hand, both \ce{HCN} and \ce{H2O} abundances (relative to \ce{CO}) differ between the TMC1A disk, inner disks as observed in the mid-IR, and comets.  However, much deeper and spatially resolved HCN and water observations are needed to fully reveal the physical processes that can cause these variations.

\subsection{Accretion processes in a young disk}
\label{sub:accretion}

\begin{table}
\caption{Reaction network used to calculate the \hcop abundance given values for \ce{H}, \ce{H2}, and CO. Reactions with a cosmic ray (CR) is also included.  Reaction rate coefficients are obtained from the Kinetic Database for Astrochemistry \citep[KiDA,][]{kida}.
}
\centering
    \begin{tabular}{c l c c}
    \hline \hline
    No. & Reaction \tablefootmark{a}         & $a$  & $b$  \\
    \hline
    1. & \ce{H2}         + CR        $\rightarrow$ \ce{H2^{+}} + \ce{e^{-}}
                & ...\tablefootmark{b}       & ...       \\
    2. & \ce{H2^{+}}     + \ce{H2}   $\rightarrow$ \ce{H3^{+}} + H
                & $2\times 10^{-9}$     & 0         \\
    3. & \ce{H3+}     + \ce{N2}   $\rightarrow$ \ce{N2H^{+}} + \ce{H2}
                & $1.70\times 10^{-9}$     & 0         \\
    4. & \ce{H3^{+}}     + \ce{CO}   $\rightarrow$ \ce{HCO^{+}} + \ce{H2}
                & $1.61\times 10^{-9}$     & 0         \\
    5. & \ce{N2H^{+}}    + \ce{CO}   $\rightarrow$ \ce{HCO^{+}} + \ce{N2}
                & $8.8 \times 10^{-10}$     & 0         \\
    6. & \ce{HCO^{+}}     + \ce{e^{-}}   $\rightarrow$ \ce{H} + \ce{CO}
                & $2.8 \times 10^{-7}$     & -0.69         \\
    7. & \ce{N2H^{+}}     + \ce{e^{-}}   $\rightarrow$ \ce{NH} + \ce{N}
                & $1.3 \times 10^{-8}$     & -0.84         \\
    8. & \ce{N2H^{+}}     + \ce{e^{-}}   $\rightarrow$ \ce{N2} + \ce{H}
                & $2.47 \times 10^{-7}$     & -0.84         \\
    9. & \ce{H3+}     + \ce{e^{-}}   $\rightarrow$ \ce{H2} + \ce{H}
                & $2.34\times 10^{-8}$      & -0.52         \\
    10. & \ce{H3+}     + \ce{e^{-}}   $\rightarrow$ \ce{H} + \ce{H} + \ce{H}
                & $4.36\times 10^{-8}$      & -0.52         \\
    \hline
    \end{tabular}
    \label{tab:chemical}
    \tablefoot{
    \tablefoottext{a}{The reaction rate coefficient is given by $k = a \times (T/300 {\rm K})^{b}$ cm$^{3}$ s$^{-1}$.  These reaction rates are valid up to 300 K.  }\tablefoottext{b}{See text for the CR ionization rate.}
    }
\end{table}

During the formation of a star and disk, mass flows from the large-scale envelope to the disk and the young protostar.  Previously, disk accretion rates have been measured through the bolometric luminosity and/or molecular emission at large radii ($> 500$ au).  On the other hand, a detailed 2D model of a bursting Class I protostar can capture the current disk structure and infer the more accurate accretion rate during the burst \citep[e.g.,][]{GBaek20,LeeYH20}.  From the bolometric luminosity of TMC1A ($L_{\rm bol} = 2.7$ $L_{\odot}$, \citealt{kristensen12}), the accretion rate is inferred to be $\sim 3\times 10^{-7}$ $M_{\odot}$ yr$^{-1}$.  Similar values are derived from molecular line observations \citep[e.g.,][]{aso15, mottram17}.  These inferred accretion rates usually refer to the mass flow from envelope-to-disk or from disk-to-star (i.e.\ stellar).  With our spatially resolved molecular line observations and associated analysis, it is possible to re-examine the accretion rate and investigate the dominant accretion mechanism in the TMC1A disk.

An accretion disk mediates the mass transfer between the large-scale envelope and the young protostar.  It does so by shifting angular momentum from the accreting the mass, which, e.g., can result in a viscously growing disc. One of the major uncertainties and sources of continued debate in understanding the evolution of disks is the physical driver of accretion.  A parameterized and constant viscosity \citep[$\alpha$,][]{SS73} is typically adopted to describe the transport of angular momentum through the disk \citep[e.g., ][]{hueso05,visser09}, but this alone does not reveal the physical driver.  In rotating magnetized disks, the magnetorotational instability \citep[MRI,][]{balbus91, balbus03} can drive accretion (and turbulence) if the disk is sufficiently ionized \citep[e.g.,][]{balbus00}.  These proposed theories rely on the kinematical structure of the disk being nearly Keplerian.  In order to constrain whether MRI is active in the TMC1A disk, which is indeed a Keplerian disk, we need to estimate the ionization fraction, and \hcop\ can be used to do this.

The abundance of \hcop\ is chemically linked to the electron abundance, $X_e$, which can be used to determine the ionization fraction.  A simple chemical model that links CO to \hcop\ and \ce{N2H^{+}} can be found in \citet[][but see also \citealt{jorgensen04c}]{aikawa15}.  Table~\ref{tab:chemical} lists the reaction network from \citet{aikawa15} that we employ here.  We adopt a fixed CO abundance of $10^{-4}$ and a \ce{N2} abundance of $3 \times 10^{-6}$ relative to \ce{H2}.  The high relative abundance of CO leads to it dominating the chemistry, and the adopted value of \ce{N2} abundance does not affect our results.  We compute the electron number density from the cosmic ray (CR) ionization rate $\zeta$ using $n_{e^{-}} = 2\sqrt{\zeta / \left ( 2 k_{6} n_{\rm H} \right )} n_{\ce{H2}}$ \citep{aikawa15} where $k_{6}$ is the rate coefficient of the sixth reaction in Table \ref{tab:chemical}.  A range of \ce{H2} densities, temperature, and $\zeta$ values are explored to investigate the effect on the resulting \hcop\ abundance.  With \ce{CO}, \ce{H2}, \ce{N2}, and \ce{e^{-}} abundances known, we solve for the equilibrium abundance of \hcop.

Using these approximations, the inferred \hcop\ abundance can be reproduced with $\zeta \sim 10^{-17}$ s$^{-1}$, \ce{H2} densities of $10^{6}$ cm$^{-3}$, and a gas temperature of 20 K.  We find that the gas density $n_{\ce{H2}}$ is the dominant factor in the calculation, while the gas temperature does not strongly affect the results; the variance in \hcop\ abundance is less than a factor of 2 for temperatures between 20 and 100 K.  From varying the \ce{H2} density, we find that \hcop\ emission seems to be confined to regions with $n_{\ce{H2}}< 10^{8}$ cm$^{-3}$ (see \citealt{mathews13}).

An MRI active region is defined by its magnetic Reynolds number \begin{equation}
    R_{\rm e} = \frac{c_{\rm s} h}{D} \approx
    1 \left ( \frac{X_{\ce{e}}}{10^{-13}} \right )
    T_{\rm 100K}^{1/2} R_{\rm au}^{3/2},
\end{equation}
where $c_{\rm s}$ is the sound speed, $h$ is the disk scale height, $D$ is the magnetic diffusivity, $T_{\rm 100K}$ is the gas temperature normalized to 100 K, and $R_{\rm au}$ is the radial distance from the star normalized to 1~au \citep{perezbecker11}.  A secondary condition for a MRI unstable disk is a high ion-neutral collision rate, i.e., that the turbulence is efficiently transferred to the bulk neutral disk.  The ion-neutral collision rate can be expressed using
\begin{equation}
    Am \approx 1 \left ( \frac{X_{i}}{10^{-8}} \right ) n_{\rm 10}
    R_{\rm au}^{3/2},
\end{equation}
where $X_{i}$ is the abundance of a singly ionized species and $n_{\rm 10}$ is the gas number density normalized to $10^{10}$ cm$^{-3}$ \citep{perezbecker11}.  Here, we assume that \hcop\ is the dominant ion.  Given the estimated abundance of the \hcop\ emitting layer, and the inferred electron abundance $X_e$ (using $n_{\ce{H2}} = 10^{6}$ cm$^{-3}$ and $\zeta = 10^{-17}$ s$^{-1}$; see above), we estimate that $R_{\rm e} > 10^{6}$ and $Am < 0.1$.  Depending on the disk structure and the magnetic field orientation, the region of the disk needs to be above the critical $R_{\rm e} = 10^2$--$10^4$ \citep{fleming00,flock12} for MRI to be active.  Magneto-hydrodynamic disk shearing box simulations by \citet{xnbai11} meanwhile suggest that MRI can be active in disks at any given $Am$ if the disk is sufficiently weakly magnetized.  Therefore, unless the TMC1A disk is weakly magnetized, which is in contrast to the magnetically-powered disk wind observed in TMC1A \citep{bjerkeli2016}, MRI is likely not active in the observed molecular layer.

The current disk accretion rate can also be inferred through the location of the water snow surface.  We use the brightness temperature profile in Fig.~\ref{fig:TMC1Atemp} to estimate that the water snow surface in TMC1A is located at $\simeq$10 au (i.e.\ where $T < 160$ K), which is consistent with the non-detection of millimeter \ce{H2^{18}O} emission from this disk \citep{harsono20}.  The relation between the midplane water snowline and the disk accretion rate in embedded disks was explored in \citet{harsono15b}.  From those results, and the non-detection of the water line, we infer that the stellar accretion rate is $\lesssim 10^{-5}$ M$_{\odot}$ yr$^{-1}$ in TMC1A.

The observed brightness temperature profile (Sect.~\ref{sub:tempstructure}) likely traces the irradiated disk surface ($T_{\rm eff} \sim L_{\star}^{1/4} R^{-1/2}$).  From the inferred water snowline location and the current bolometric luminosity, we estimate that the current stellar accretion rate is close to $10^{-6}$ M$_{\odot}$ yr$^{-1}$.  Note that the current TMC1A disk accretion rate ($\sim 10^{-6}$ $M_{\odot}$ yr$^{-1}$) is higher than the values obtained from its bolometric luminosity and previous large-scale molecular emission.  A more detailed 2D physical structure \citep[see e.g.,][]{cleeves13,cleeves17} of the disk+envelope is, however, required to more accurately assess these values.

Other sources of angular momentum transport which could drive the accretion in the TMC1A disk could be magnetically-powered winds \citep[e.g.][]{xnbai13, ramsey2019}, and gravitational (GI) or hydrodynamical instabilities \citep{lyra19}.  Due to the absence of observable dust substructures in TMC1A, \citet{harsono18} suggest that $M_{\rm disk}/M_{\star} \lesssim 0.1$, yielding $\alpha_{\rm GI} \lesssim 0.06$ (i.e.\ a long cooling timescale $t_{\rm cool} \sim 10 \Omega^{-1}$).

%
%
\section{Summary and Conclusions}
\label{sec:summaryconclude}

This paper presents spatially resolved observations of dust and gas with ALMA toward the young disk around TMC1A.  The high-spatial resolution provided by 16 km baselines has proven crucial in isolating the emission of the young disk from its surrounding envelope.  Studies such as this are critical to tracing the evolution of various molecules during the early stages of planet formation.  The results and conclusions of this paper are as follows.
\begin{itemize}

    \item The dust disk is detected at 203, 220, 230, 240, and 260 GHz.  Dust continuum visibilities are analyzed with Gaussian intensity profiles at each frequency to constrain the orientation of the disk around TMC1A to $i=50^{\circ} \pm 3^{\circ}$ and $PA = 76^{\circ} \pm 4^{\circ}$.

    \item We present high-spatial observations of \ce{DCN} 3--2, \ce{HCN} \mbox{3--2}, \hcop\ 3--2, and SO $5_6$--$4_5$, as well as a hint of spectrally unresolved \ce{N2D^{+}} emission.  The \ce{N2D^{+}} emission cannot be imaged because it is mostly filtered out.

    \item High-spatial-resolution CO observations are essential to distinguish the molecular emission associated with the disk wind, the envelope, and the Keplerian disk.  By comparing the morphology of the \hcop\ emission to CO, we determine that \hcop\ traces the upper layers of the disk and parts of the infalling envelope.

    \item Two HCN emission peaks are located to the west on the red-shifted side of the disk and to the north of the blue-shifted side of the disk.  By comparing the HCN to \mco\ and \ceo, the narrow line profile suggests that the emission to the north of the disk traces the protostellar envelope near the outflow cavity wall. Meanwhile, the red-shifted HCN emission to the west is co-spatial with the \hcop\ emission and emanates from the surface of the embedded Keplerian disk.

    \item The zeroth moment map of SO shows a partial ring-like structure that has been seen towards other protostellar systems.  Owing to the orientation of the TMC1A system, we are able to differentiate between the plane of SO emission and the Keplerian disk as traced by \ceo.  The combination of the SO line profile and its low brightness temperature indicates that the emission originates from the dense and warm inner envelope close to the outflow cavity wall.

    \item The molecular emission is analyzed considering thermalized, optically thin emission as well as non-LTE models using the \textsc{RADEX} radiative transfer code.  We find that the optically thin limit provides a reasonable estimate of the emitting column densities of the detected molecular lines.  With the derived column densities, we infer the abundance structure of the disk relative to CO.  The disk-averaged abundances are then $X_{\ce{HCO^{+}}} = 10^{-10}$\,--\,$10^{-9}$, $X_{\ce{HCN}} = 10^{-11}$\,--\,$10^{-9}$ and $X_{\ce{SO}} = 10^{-9}$\,--\,$10^{-7}$.  With an upper limit to the DCN emission, we estimate a \ce{DCN}/\ce{HCN} ratio of $<0.26$ for the TMC1A disk.

    \item The comparison of molecular abundances of the disk to its surrounding envelope shows that the observed molecular layer of the disk is composed of unaltered infalling gas.  This indicates a smooth transition between the envelope and the young disk.  Furthermore, the similarity of the relative abundances of \hcop, HCN, and SO between the young TMC1A disk and Class II disks suggest that some parts of the molecular layer in disks are set during the disk formation phase.

    \item The derived HCN, DCN, and \ce{H2O} molecular abundances of the TMC1A disk show larger discrepancies relative to Class II disks and Solar System objects (comet 67P, Hale Bopp, and Titan).  While the \ce{HCN}/\ce{H2O} ratio of the TMC1A disk is consistent with observed inner disks and comets, the ratio is different from values typically found in protostellar envelopes.  Similarly, the individual \ce{HCN} abundance of the TMC1A disk is different from Solar System comets despite that it is within the range of other Class II disks.  From these comparisons, it would seem that the ratio of \ce{HCN}/\ce{H2O} is established during the disk formation process.  We propose that deeper observations of HCN isotopologs and \ce{H2O} are crucial to understand the early physical and chemical evolution of planet-forming disks.

    \item
    Explaining the accretion process in disks is one of the fundamental problems in astrophysics.  With the derived \hcop\ abundance, we find that the observed molecular layer of the TMC1A disk is not sufficiently ionized to be MRI unstable.  The ionization rate is obtained using a reduced chemical network to reproduce the observed \hcop\ abundance.  We estimate an accretion rate of the TMC1A disk of $\sim 10^{-6}$ $M_{\odot}$ yr$^{-1}$.  Other physical processes such as disk winds, gravitational instability, or hydrodynamical instabilities are thus needed to drive accretion in TMC1A.

\end{itemize}

These results are one of the first that directly compare the relative molecular abundances (six species) in a young disk with its own envelope, Class II disks, and Solar System objects.  The aim of this comparison is to provide molecular tracers that can probe the physics of disk formation and accretion.  In addition, \hcop\ and \ce{H2O} observations are crucial in revealing the accretion process (envelope-to-disk, disk-to-star) during the embedded stage of star formation.  These observations support the idea that the composition of the molecular layer of planet-forming disks is determined during the disk formation phase.  Future deep observations that require ALMA LBC ($> 16$ km baselines) per target ($L_{\rm bol} \sim 3 L_{\odot}$) within 200 pc will be needed to further unravel the chemical structure of Keplerian disks around young stellar objects.

\begin{acknowledgements}
This paper makes use of the following ALMA data: ADS/JAO.ALMA\#2015.1.01549.S, ADS/JAO.ALMA\#2016.1.00711.S, and ADS/JAO.ALMA\#2017.1.00212.S.  ALMA is a partnership of ESO (representing its member states), NSF (USA) and NINS (Japan), together with NRC (Canada) and NSC and ASIAA (Taiwan) and KASI (Republic of Korea), in cooperation with the Republic of Chile.  The Joint ALMA Observatory is operated by ESO, AUI/NRAO and NAOJ.
This work uses observations carried out under project number V064 and X065 with the IRAM NOEMA Interferometer/PdBI. IRAM is supported by INSU/CNRS (France), MPG (Germany) and IGN (Spain).
This work is based on observations collected at the European Southern Observatory Very Large Telescope under program ID 179.C-0151.  We thank Klaus Pontoppidan and Greg Herczeg for providing the calibrated data.
We thank the anonymous referee for carefully reading the manuscript and their constructive feedback that improved this paper.
Astrochemistry in Leiden is supported by the European Union A-ERC grant 291141 CHEMPLAN, by the Netherlands Research School for Astronomy (NOVA) and by a Royal Netherlands Academy of Arts and Sciences (KNAW) professor prize.  D.H.\ acknowledges support from the EACOA Fellowship from  the  East  Asian Core  Observatories Association.  P.B.\ acknowledges the support of the Swedish Research Council (VR) through contracts 2013-00472 and 2017-04924. J.P.R.\ acknowledges support from Virginia Initiative on Cosmic Origins (VICO), and the National Science Foundation (NSF) under grant nos.\ AST-1910106 and AST-1910675.  J.K.J.\ acknowledges support by the European Research Council (ERC) under the European Union's Horizon 2020 research and innovation programme through Cosolidator Grant ``S4F" (grant agreement No~646908).
This research made use of Astropy community-developed Python package for Astronomy \citep{astropy18}, numpy, matplotlib \citep{Hunter2007}, and python package \textsc{casacore} to handle CASA products (images and measurement sets).

\end{acknowledgements}


\bibliographystyle{aa} 
\bibliography{biblio.bib} 

\begin{appendix} 

\section{Undetected molecular lines}
\label{app:undetected}

\begin{table*}
    \caption{Undetected molecular lines in our spectral windows.  Synthesized beams and noise levels per velocity channel (0.3 km s$^{-1}$) are shown. }
    \label{tbl:upperlims}
    \centering
    \begin{tabular}{lccccc}
        \hline \hline
        Molecular line transition    &   Frequency   & $E_{\rm up}$
        & $\log_{10} A_{\rm ul}$  & Beam\tablefootmark{a} &  Noise
        \\
                    & (GHz)             & (K)     &  (s$^{-1}$)
                & ($'' \times '', ^{\circ}$)      & (mJy bm$^{-1}$)
        \\
        \hline
       \ce{SiO}  $J=5-4$    & 217.10450      & 31.26  & $-3.28$
        & $0.13 \times 0.10$, 32                 & 1.6
        \\
       \ce{^{13}CN} $N_J = 2_{3/2} - 1_{1/2}, F_1,F = 1,1-0,1$
       & 217.27768      & 10.88     & $-4.24$
        & $0.08 \times 0.07$, 78                 & $\sim5$
        \\
       \ce{^{13}CS}  $J=5 - 4$    & 231.22069  & 33.29  & $-3.60$
        & $0.12 \times 0.10$, 32                 & 1
        \\
       \ce{SO}  $N_J=3_4 - 4_3$    & 267.19774  & 19.93  & $-6.15$
        & $0.13 \times 0.10$, 3                 & 1.3
        \\
        \hline
    \end{tabular}
\tablefoot{
  \tablefoottext{a}{Elliptical synthesized beam parametrized by: FWHM long axis $\times$ FWHM short axis, position angle of the long axis. }
    }
\end{table*}

There are additional molecular lines present in our ALMA spectral set ups that are not detected in our data.  Table~\ref{tbl:upperlims} lists these lines. The \ce{^{13}CN} line is weakly apparent in the visibilities and could be identified in a spectrum taken over a large area ($> 10$ beams).  However, it cannot be imaged properly even including some tapering.  The noise level per channel is higher than the other lines because low-level \ce{^{13}CN} permeates the spectral cube.  This implies that most of the emission is filtered out in our long baseline data.  SO $N_{J} = 3_4 - 4_3$ is likely not detected because of its low Einstein A coefficient; for the physical conditions that produce the detected SO $5_6 - 4_5$ line, the strength of the $3_4 - 4_3$ line would be a factor of 100 weaker.

\section{Molecular lines: vibrational transitions}
\label{app:vibrational}

\begin{figure}
 \centering
 \includegraphics[width=0.48\textwidth]{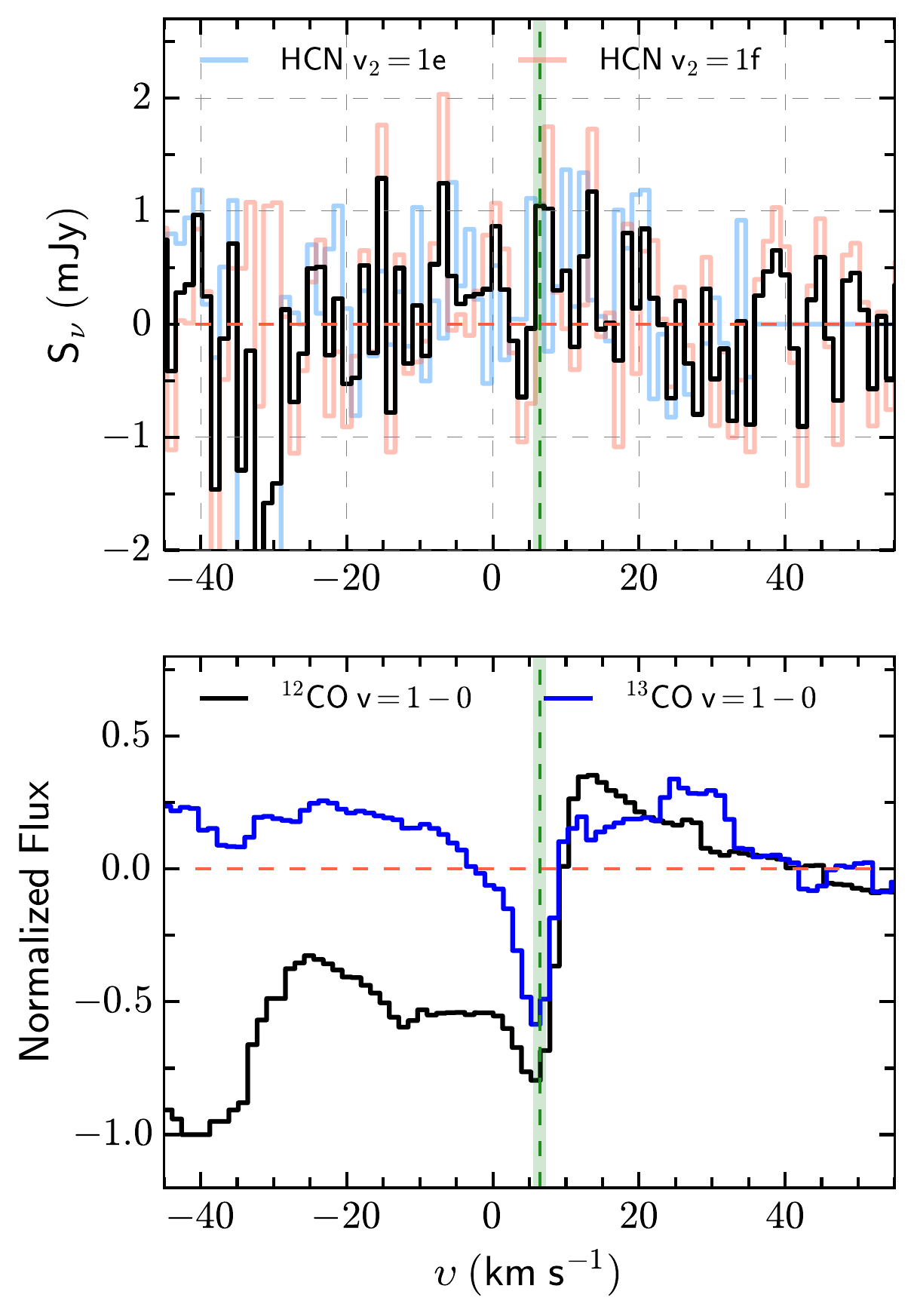}
 \caption{
 {\it Top}: HCN rovibrational spectra ($v_2 = 1$ $J = 3-2$ $e/f$) observed with ALMA.  The spectra are extracted from the inner 1$''$ region with red and blue denoting the $e$ and $f$ vibronic angular momentum quantum number, respectively.  The black spectrum shows the stacked spectrum of the two lines, allowing a more stringent upper limit on the transition.  {\it Bottom}: Fundamental $v=1-0$ CO vibrational spectra taken with the VLT/CRIRES \citep{herczeg11}.  The shown \mco\ spectrum is the co-added rotational line profiles from R(3) to P(9), while the \tco\ spectrum includes co-added R(10) to P(9).  The green vertical solid line is the systemic velocity of the system, while the red horizontal dashed line indicates the baseline.
 }
 \label{fig:rovibspectra}
\end{figure}

Our spectral cubes also contain the HCN bending mode ($\upsilon_2 = 1$) rotational transitions at 265.8527 GHz and 267.1993 GHz, as shown in Fig.~\ref{fig:rovibspectra}. These lines are not detected toward TMC1A. They do, however, provide a constraint on the physical structure of TMC1A disk due to their high upper energy level ($E_{\rm up} \sim 1000$ K).  The peak flux density of each line has a 0.9 mJy upper limit.  For comparison, we also present the fundamental vibrational CO lines taken with CRIRES on the VLT \citep{herczeg11} in the same figure with upper energy levels $\sim 3000$ K.  These spectra are taken with a 0.2$\arcsec$ slit at a $PA$ of 64$^{\circ}$, which is along the major axis of the Keplerian disk.  A deep absorption is apparent at the systemic velocity of the coadded \mco\ $\upsilon=1-0$ and \tco\ $v=1-0$ spectra, and is similar to what we see in the rotational spectra in the ground vibrational state (Fig.~\ref{fig:rotspectra}). The broad blue absorption in the CO fundamental vibrational line is, meanwhile, a strong indication of a disk wind \citep{calvet92,pontoppidan11, bast11,herczeg11}.

In order to further constrain the HCN $\upsilon_2=1$ emission, the weighted average of the two spectra ($e$,$f$) is also shown in black in Fig.~\ref{fig:rovibspectra}.  The stacked spectrum shows tantalizing features of beam diluted vibrational HCN emission. We expect that the emission should follow the \tco\ $\upsilon=1-0$ line profile as the red-shifted emission ($> 8$ km s$^{-1}$) comes from the hot surface layer of the disk instead of the disk wind \citep{herczeg11}.  Therefore, we believe that the absorption feature in the stacked HCN spectrum and emission near the systemic velocity could be firmly detected with deeper observations.

\end{appendix}
%
%
\end{document}